\documentclass[reprint,longbibliography,amsmath,amssymb,prd,superscriptaddress]{revtex4-2}
\usepackage[utf8]{inputenc}
\usepackage[T2A]{fontenc}
\usepackage[russian,main=english]{babel}
 \usepackage{relsize}
\usepackage{graphicx}
\usepackage{dcolumn}
\usepackage{bm}
\usepackage[colorlinks=true, citecolor=blue, linkcolor=blue, urlcolor=blue]{hyperref}
\usepackage{xcolor}
\usepackage{amsmath, amssymb, amsfonts}
\usepackage{mathtools}
\usepackage{subfigure}
\usepackage{mathrsfs}
\usepackage{sidecap}
\usepackage{verbatim}
\usepackage{multirow}
\usepackage{amsthm}
\usepackage{tikz}
\usetikzlibrary{matrix}
\theoremstyle{definition}

\definecolor{purple1}{rgb}{128,0,128}

\newcommand{\nn}{\nonumber\\}
\usepackage{bigints}
\newcommand{\bea}{\begin{eqnarray}}
\newcommand{\ea}{\end{eqnarray}}

\definecolor{darkpastelgreen}{rgb}{0.01, 0.75, 0.24}
\makeatletter
\newcommand*{\rom}[1]{\expandafter\@slowromancap\romannumeral #1@}
\makeatother
\begin{document}
\title{Petrov classification of analogue spacetimes}
\author{Sang-Shin Baak} 
\author{Satadal Datta}
\author{Uwe R. Fischer}
\affiliation{%
Seoul National University, Department of Physics and Astronomy, Center for Theoretical Physics, Seoul 08826, Korea}
\date{\today}

\begin{abstract}
In an effort to invariantly characterize the conformal curvature structure   
of analogue spacetimes built from a nonrelativistic fluid background, 
we determine the Petrov type of a variety of laboratory geometries.   
Starting from the simplest examples, we increase the complexity of the background, and thereby 
determine how the laboratory fluid symmetry affects the corresponding Petrov type in the analogue spacetime 
realm of the sound waves. We find that for more complex flows isolated hypersurfaces develop,  
which are of a Petrov type differing from that of the surrounding fluid. 
{Finally, we demonstrate that within the incompressible background approximation, as well as for all
compressible quasi-one-dimensional flows, the only possible Petrov types are the algebraically general type I 
and the algebraically special types O and D.} 

\end{abstract}

\maketitle
\section{Introduction}
A pioneering idea of Unruh  more than four decades ago \cite{unruh}, with an early precursor by Trautman \cite{Trautman},  
created what is now termed {\em analogue gravity} \cite{BLV}:  In an inviscid irrotational barotropic fluid, a linearized 
sound perturbation propagates like a massless minimally coupled Klein-Gordon field in an effective spacetime. 
The corresponding acoustic spacetime metric is determined by  the background flow as a solution of the fluid equations under the influence of some external potential, with possibly also an additionally  
engineered equation of state, which can be created in the lab. Considering the nonrelativistic fluid equations \cite{Landau1987Fluid} 
in a Newtonian inertial frame, 
Unruh demonstrated theoretically that 
an acoustic spacetime mimicking a black hole spacetime 
with an event horizon can be designed by a background flow speed which exceeds the 
speed of sound of the medium at some radius, for a spherically symmetric steady state flow. 
The quantization of the linearized perturbation 
then yields an experimentally detectable Hawking radiation from such a ``dumb'' (or sonic) black hole. 
The resulting possibility of an experimental verification of 
quantum field theories in {\em fixed} curved spacetime backgrounds 
(that is, as described by the background field method) is an enticing prospect.  
As a result, with recent advances in experimental capabilities in particular for quantum gases, the 
Hawking radiation effect for sonic black holes   \cite{PhysRevLett.105.240401}  
has been observed in a Bose-Einstein Condensate \cite{Munoz,Steinhauer2021}, 
with further intriguing prospects of an interplay of theory and experiment 
opening up also, e.g., in the context of quantum cosmology 
\cite{Grisha,Fedichev,Schuetzhold,unruh2007quantum,PhysRevLett.118.130404,Chin,Eckel,Viermann}. 


In what follows, we probe deeper into the classification of analogue gravity spacetimes, by investigating the algebraic properties of the Weyl tensor \cite{Weyl}, 
via the Petrov classification, which has been first laid down in Ref.~\cite{petrovRussian} (an English translation is found in 
\cite{Petrov2000}, also see the monograph \cite{petrovbook}). This method  
has also been referred to as Petrov-Penrose or Petrov-Pirani-Penrose classification due to the important contributions made in Refs.~\cite{Pirani1957,npf,Penrose1986}. 
One major motivation of using this classification scheme in the present analogue context is the possibility of 
an invariant characterization of a given spacetime by means of its conformal curvature structure. This, then, indicates the degree to which analogue and Einsteinian spacetimes can be similarly classified with respect to their conformal structure, 
which is an aspect separate from the (very different) dynamical origin of the spacetimes. 
{Thus the Petrov classification of a given spacetime metric, which is kinematical in nature,}
 places further emphasis on the importance of differentiating essentially 
kinematical from dynamical aspects of a curved spacetime. The latter distinction of kinematical versus dynamical 
is salient also, e.g., for properly formulating the requirements to observe Hawking radiation and black hole entropy, respectively \cite{VisserPRL1998}. 

\section{Riemann and Weyl tensors}
\subsection{General relation}
Using Riemann normal coordinates, any metric $g_{\mu\nu}$ 
locally can be rendered equal to the Minkowski metric $\eta_{\mu\nu}$, 
and all of its first order derivatives $\partial_\kappa g_{\mu\nu}$ vanish.
In a $D$ dimensional spacetime manifold, we have $\frac14 D^2(D+1)^2$ independent 
second derivatives  of $\partial_\lambda\partial_\kappa g_{\mu\nu}$. 
These second-order derivatives can in a curved manifold 
never be rendered all zero by 
suitable local coordinate transformations. 
There are always $\frac1{12} D^2 (D^2 -1)$ independent nonzero second-order derivatives remaining, which thus contain the information about the {\em curvature} of a manifold.  The Riemann tensor $R_{\mu\nu\lambda\kappa}$ \cite{Riemann1921} involves these second-order derivatives, and possesses the required symmetries. Hence it has the required 
$\frac1{12} {D^2}(D^2 -1)$ independent components, and defines the general curvature of a manifold.  
The only second rank covariant tensor which can be constructed from the Riemann tensor by contraction is the symmetric Ricci tensor, $R_{\mu\nu}=g^{\lambda\kappa}R_{\lambda\mu\kappa\nu}$, and further contraction of indices give us the Ricci scalar $R=g^{\mu\nu}R_{\mu\nu}$, where 
$g^{\mu\kappa}g_{\kappa\nu}=\delta ^{\mu}{}_{\nu}$. The number of independent $R_{\mu\nu}$, $\frac12 D(D+1)$,  
is identical to number of  independent $g_{\mu\nu}$. Therefore, only in spacetime dimension $D\geq 4$, the full Riemann tensor is needed to describe the curvature of spacetime. In $D\geq 4$, it has additional $\frac{1}{12}D^2(D^2 -1)-\frac12 D(D+1)=\frac{1}{12}D(D+1)(D+2)(D-3)$ independent components over those of $R_{\mu\nu}$. 
One can then introduce the Weyl tensor $C_{\mu\nu\lambda\kappa}$, which is representing these additional independent components (we are following the conventions of Ref.~\cite{weinberg1972gravitation}), 
as follows 
\begin{multline}\label{Rt}
C_{\mu\nu\lambda\kappa}=
R_{\mu\nu\lambda\kappa}\\
-\frac{1}{D-2}\left(g_{\mu\lambda}R_{\nu\kappa}-g_{\mu\kappa}R_{\nu\lambda}-g_{\nu\lambda}R_{\mu\kappa}+g_{\nu\kappa}R_{\mu\lambda}\right)\\
+\frac{R}{(D-1)(D-2)}\left(g_{\mu\lambda}g_{\nu\kappa}-g_{\mu\kappa}g_{\nu\lambda}\right).
\end{multline}
\subsection{Curvature in acoustic spacetimes} 
The expression \eqref{Rt} tells us that the curvature tensor can be written in terms of a Weyl tensor part and a part which depends on the Ricci tensor and the metric tensor.  For a spacetime which satisfies the vacuum Einstein equations, $R_{\mu\nu}=0$, $R=0$; therefore, Eq.~\eqref{Rt} implies 
$R_{\mu\nu\lambda\kappa}=C_{\mu\nu\lambda\kappa}$. 
We note however that since the acoustic metric 
is determined by the background flow, derived from the fluid equations, the nontrivial cases 
of acoustic spacetimes discussed below 
do not necessarily satisfy the vacuum Einstein equations. 

The Riemann curvature tensor of an acoustic spacetime manifold manifests itself through the geodesic deviation of sound rays (in the geometrical acoustics limit), propagating on a given background flow \cite{FISCHER200322}, and one may study as an application for example the gravitational lensing of sound caused by an irrotational vortex flow \cite{iv}. 
The geometric acoustic metric 
determines the path of a large momentum sound ray in the medium, and can be  simply defined to be the physical acoustic metric with the conformal factor set to unity \cite{BLV}. 
This, then, addresses the conformal part of the Riemann curvature tensor, i.e., 
the Weyl tensor part of Eq.~\eqref{Rt}. 

\section{Petrov classification}
To make our presentation sufficiently self-contained, and accessible to a wide range of communities, 
we now discuss the algebraic essentials of the Petrov classification scheme we employ.

\subsection{Null tetrads and Weyl scalars}\label{nullandWeyl}
A given spacetime geometry can be classified in terms of algebraic properties of the Weyl tensor by 
the Petrov classification scheme 
\cite{petrovRussian,petrovbook,d1992introducing}. 
Curvature is a local property of spacetime, therefore the Petrov type determines the local algebraic properties of the spacetime geometry. 

{To be more specific, in the Newman-Penrose formalism 
\cite{npf,stephani2009exact},  
here in $D=4$ for a pseudo-Riemannian manifold, one works with a specific ``null''  choice of the tetrad basis instead of the more usual orthonormal tetrad basis. 
As a simple example, flat spacetime 
is represented in the Newman-Penrose formalism by  lightcone coordinates, i.e., 
one introduces the following complex linear coordinate transformation (here we put 
speed of light (sound) $c=1$) }
\begin{eqnarray}
& u=\frac{1}{\sqrt{2}}(t-x), \quad 
v= \frac{1}{\sqrt{2}}(t+x),\\
& w= \frac{1}{\sqrt{2}}(y+{i}z), \quad 
\bar{w}=\frac{1}{\sqrt{2}}(y-{i}z). \label{Mink}
\end{eqnarray}
By convention, we use lower case Latin letters $a,~b,,..$ to denote tetrad indices, 
and Greek letters $\mu,\nu,...$ denote indices in the coordinate basis.
With these new, complex-valued coordinates $(u,v,w,\bar{w})$, the Minkowski metric is  
\begin{eqnarray}\label{teta}
\eta_{ab}=\eta^{ab} \coloneqq \begin{bmatrix}
 0 & -1 & 0 & 0\\
-1 & 0 & 0 & 0 \\
0 & 0 & 0 & 1\\
0 & 0 & 1 & 0
\end{bmatrix}. 
\end{eqnarray}

{In general, a metric tensor is expressed by using 
the tetrad basis as follows}  \footnote{We remark that for any tetrad basis variant, orthonormal or null, 
{there are remaining tetrad gauge degrees of freedom},  
which can be used to bring the tetrad into a suitable form for a specific computation.}
\begin{equation}\label{gmnt}
g_{\mu\nu}=e_{\mu a}e_{\nu b}\eta ^{ab}.
\end{equation}
Tetrad indices are raised or lowered by the Minkowski metric of \eqref{teta}, 
and the spacetime metric (and its inverse) lowers (and raises) Greek indices. 
Therefore, using the identity $g^{\mu\sigma}g_{\sigma\nu}=\delta^{\mu}_{~\nu}$, and Eq.~\eqref{gmnt}, we have $e_{\nu a}e^{\mu a}=\delta^{\mu}_{~\nu}$, and hence
\begin{equation}\label{tc}
\eta _{ab}=e^{\mu}_{~a}e^{\nu}_{~b}g_{\mu\nu}=e_{\mu a}e^{\mu}_{~ b}.
\end{equation}
Expressing the  flat spacetime metric in this way 
leads to two real and two complex null tetrad vectors. 
Following the conventions of Ref.~\cite{npf}, 
we denote {the real null tetrad basis vector {components} by} $e_{\mu 0}= l_\mu$, $e_{\mu 1}= n_\mu$, 
{and the complex null tetrad basis vectors by} 
$e_{\mu 2}= m_\mu$, and $e_{\mu 3}= \bar{m}_\mu$. Here, $\bar{m}_\mu$ is the complex conjugate of $m_\mu$. 
{In Minkowski space, upon aligning the null tetrad basis to the coordinate basis vectors provided by the coordinates in \eqref{Mink}, we have 
$ \partial _u=l, \,
    \partial_v =n, \,
    \partial_w=m,\, \mbox{and}\,  
    \partial_{\bar{w}}=\bar{m}$. }

{For a general metric $g_{\mu\nu}$, Eq.~\eqref{tc} provides us with}   
\begin{eqnarray}
l_\mu l^\mu=n_\mu n^\mu=m_\mu m^\mu=\bar{m}_\mu \bar{m}^\mu=0,\label{ll}\\
l_\mu n^\mu= -m_\mu \bar{m}^\mu=-1, \label{ln}\\
l_\mu m^\mu= l_\mu \bar{m}^\mu=n_\mu m^\mu= n_\mu \bar{m}^\mu=0. \label{lm}
\end{eqnarray}
Then, we have from Eq.~\eqref{gmnt}, 
\begin{equation}
g_{\mu\nu}=-l_\mu n_{\nu}-n_{\mu} l_{\nu}+m_{\mu}\bar{m}_{\nu}+\bar{m}_{\mu}m_{\nu}.
\end{equation}
In spacetime dimension $D=4$, the 10 independent Weyl tensor components can be represented in terms of 5 complex scalars. 
Using the tetrad components $C_{abcd}=C_{\lambda\mu\nu\kappa}e^{\mu}_{~a}e^{\mu}_{~b}e^{\mu}_{~c}e^{\mu}_{~d}$, the five Weyl 
scalars are defined as follows \cite{npf}
\begin{align}
 \Psi_0 & \coloneqq C_{0202}=C_{\lambda\mu\nu\kappa}l^{\lambda}m^{\mu}l^{\nu}m^{\kappa},\nn
 \Psi_1  & \coloneqq C_{0102}=C_{\lambda\mu\nu\kappa}l^{\lambda}n^{\mu}l^{\nu}m^{\kappa},\nn
\Psi_2  & \coloneqq  C_{0231}=C_{\lambda\mu\nu\kappa}l^{\lambda}{m}^{\mu}\bar{m}^{\nu}n^{\kappa}
,\\
 \Psi_3  & \coloneqq C_{0131}=C_{\lambda\mu\nu\kappa}l^{\lambda}n^{\mu}\bar{m}^{\nu}n^{\kappa},\nn
 \Psi_4  &\coloneqq C_{0313}= {C_{\lambda\mu\nu\kappa}n^{\lambda}\bar{m}^{\mu}n^{\nu}\bar{m}^{\kappa}.}
 \nonumber
\end{align}

{If one can render $\Psi_0=\Psi_1=0$ by a suitable choice of tetrad basis, the spacetime is called 
algebraically special \cite{npf,stephani2009exact}.
If this is not possible, the spacetime is called algebraically general, and classified as type I.

Given $\Psi_0=\Psi_1=0$, the algebraically special spacetimes are Petrov-classified as follows:}

\begin{enumerate}
\item $\Psi_2$, $\Psi_3$ and $\Psi_4$ are nonzero: Petrov type II. 
\item Only $\Psi_3$ and $\Psi_4$ are nonzero: Petrov type III.
\item Only $\Psi_2$ is nonzero,: Petrov type D. 
\item Only $\Psi_4$ nonzero: Petrov type N.
\item  All Weyl scalars are zero (Weyl tensor is identically zero): Petrov type O.
 \end{enumerate}

\subsection{The ${\mathcal Q}$ matrix and its Segre characteristic} 
\label{QS} 

{For any choice of tetrad, one can construct from the Weyl scalars  
 a symmetric traceless matrix}  \cite{stephani2009exact} 
\begin{equation}
   {\mathcal Q} = \! \begin{bmatrix}
        \Psi_2-\frac{1}{2}(\Psi_0+\Psi_4) & \frac{i}{2}(\Psi_4-\Psi_0) & \Psi_1-\Psi_3\\
        \frac{i}{2}(\Psi_4-\Psi_0) & \Psi_2+\frac{1}{2}(\Psi_0+\Psi_4) & i(\Psi_1+\Psi_3)\\
        \Psi_1-\Psi_3 & i(\Psi_1+\Psi_3) & -2\Psi_2
    \end{bmatrix}\!.
    \label{Qmatrix} 
\end{equation}
The algebraic properties of the ${\mathcal Q}$ matrix determine the Petrov type, as summarized in table \ref{Qtable}. 
{We here choose to work with the ${\mathcal Q}$ matrix to find the Petrov type. 
Alternatively, one can induce the Petrov type by calculating the 
principal null directions {of the Weyl tensor}, by solving a quartic equation \cite{chandrasekharBH}.}

{Under a conformal transformation, the Petrov type is invariant, as follows from its 
definition via the Weyl scalars. 
Namely, one has by the conformal transformation $g'_{\mu\nu}\rightarrow \Omega ^2 g_{\mu\nu}$, with 
$\Omega$ a function of space and time, the required invariance of 
${C'}^{\alpha}_{~\beta\gamma\delta} \rightarrow C^{\alpha}_{~\beta\gamma\delta}$, while the Weyl scalars change as 
$\Psi'_n\rightarrow \frac{1}{\Omega ^2}\Psi_n ,\, n=\{0,\ldots,4\}$.}

{{\em Segre characteristic.} Given any square matrix $A$, 
one can perform a similarity transformation to make it ``as nearly diagonal as possible.'' 
More formally, $\forall A\in\mathrm{Mat}(n,\mathbb{K}) : \exists\, M\in\mathrm{GL}(n,\mathbb{K})$ such that \footnote{
{Here, $\mathrm{GL}(n,\mathbb{K})$  is the General Linear group associated to a $n$ by $n$ matrix,
 with $\mathbb K$ the mathematical {\em field} ($\mathbb{C}$ for the ${\mathcal Q}$ matrix).}}}
\begin{equation} 
    J(A) = M^{-1}AM =\begin{bmatrix}
        J_1 &  \\ 
            & \ddots &\\
            & & J_m
    \end{bmatrix}, \qquad m\leq n
\end{equation}
The resulting matrix $J$ is called Jordan normal form where each submatrix $J_i, i\in\{1,\ldots, m\}$ takes the form 
\begin{equation}
J_i =
\begin{tikzpicture}[baseline=(current bounding box.center)]
\matrix (m) [matrix of math nodes,nodes in empty cells,right delimiter={]},left delimiter={[} ]{
        \lambda_i & 1 & & \\
                  &  & & \\
                  &  & & 1\\
            & & & \lambda_i\\
} ;
\draw[loosely dotted,thick] (m-1-1)-- (m-4-4);
\draw[loosely dotted,thick] (m-1-2)-- (m-3-4);
\end{tikzpicture}
\label{eq:eqq1}
\end{equation}
where $\lambda_i$ are the eigenvalues of $A$. 
{Here and in $J(A)$, entries not shown are all zero, and the diagonal dots indicate repetition of the entry 
along the corresponding diagonal.}

\begin{table}[t] 
\begin{center}
\vspace*{1em}
\begin{tabular}
{|c|c|c|}
\hline
\,Petrov \,& \,Segre characteristic\,& \,Annihilating polynomial\,
\\\hline
 I  &  $[111]$  & $({\mathcal Q}-\lambda _1 \mathbb{I})({\mathcal Q}-\lambda _2 \mathbb{I})({\mathcal Q}-\lambda _3 \mathbb{I})$ \\
 \hline
 II &   $[21]$  & $({\mathcal Q}+\frac 12 \lambda \mathbb{I})^2({\mathcal Q}-\lambda \mathbb{I})$ \\
\hline
 III & $[3]$ & ${\mathcal Q}^3$ \\
 \hline 
 D &   $[(11)1]$  & $({\mathcal Q}+\frac 12 \lambda \mathbb{I})({\mathcal Q}-\lambda \mathbb{I})$ \\
 \hline
 N & $[(21)]$ & ${\mathcal Q}^2$ \\
 \hline
 O & & ${\mathcal Q}$\\
 \hline
\end{tabular}
\caption{The Jordan normal form of the ${\mathcal Q}$ matrix 
determines the Petrov type given in the first column via 
the Segre characteristic \cite{MooreJohn},  as indicated in the second column. 
The Petrov type is equivalently given by the (minimal)   
annihilating polynomial 
\cite{Shilov}, shown  in the third column. }
\label{Qtable}
\end{center}
\end{table}

{The Segre characteristic (or Segre symbol) is a set of positive integers with brackets which indicate the structure of the Jordan normal form of the matrix. 
The integers in the Segre characteristic denote the size of the Jordan block. If there is more than one Jordan block which has the same value of the diagonal entry, i.e., if $\lambda_i=\lambda_j$ for some $i,j\in\{1,\ldots,m\}$, the integers denoting the multiplicity of the  $J_i$ and $J_j$ are put in round brackets, enclosed by an overall square bracket 
in the Segre characteristic notation.} 

{
Since ${\mathcal Q}$ is a traceless $3\times3$ square matrix, there are three possibilities for the set containing the 
$\lambda_i$: 
\begin{itemize}
    \item[$\diamond$] $\lambda_1\neq\lambda_2\neq\lambda_3$. For this case, each Jordan matrix is just one by one matrix ($J_i=\lambda_i$).
    Hence, the Segre characteristic is $[111]$, and the spacetime is algebraically general (Petrov type I).
    \item[$\diamond$] If two of the eigenvalues are equal, say $\lambda_1=\lambda_2$, then tracelessness gives  
    $2\lambda_1=2\lambda_2=-\lambda_3$. 
    For this case, there are two subclasses 
    \begin{itemize}
        \item Each Jordan matrix is just a one by one matrix ($J_i=\lambda_i$).
        Hence, the Segre characteristic is $[(11)1]$, and we have Petrov type D.
        \item $J_1=\begin{bmatrix}
            \lambda_1 & 1\\
            0 & 
            \lambda_1 
        \end{bmatrix}$ and $J_2=\lambda_3$ so that the Segre characteristic is $[21]$ and the Petrov type is II.
    \end{itemize}
    \item[$\diamond$] $\lambda_1=\lambda_2=\lambda_3=0$.
    \begin{itemize}
        \item Each Jordan matrix is a one by one matrix ($J_i=\lambda_i=0$), a simple number.  
        In this case, $J=0$. Equivalently, ${\mathcal Q}=0$ and a  Segre characteristic is therefore not attributed.
        \item $J_1=\begin{bmatrix}
            0 & 1\\
            0 & 0
        \end{bmatrix}$ and $J_2=0$ so that the Segre characteristic is $[(21)]$ (type N).
        \item $J=J_1=\begin{bmatrix}
            0 & 1 & 0\\
            0 & 0 & 1\\
            0 & 0 & 0
        \end{bmatrix}$. Hence, the Segre characteristic is $[3]$ (type III). 
    \end{itemize}
\end{itemize}}
{{\em Annihilating polynomial.} In addition to the Segre characteristic, in 
the last column of Table \ref{Qtable}, we display the minimal annihilating polynomial corresponding to the Segre characteristic, which by definition is the polynomial of ${\mathcal Q}$ of the lowest degree which is a zero matrix. }

{\em Examples from Einstein gravity.} 
{The very simplest cases are the Minkowski and Friedmann–Lema\^\i tre–Robertson–Walker metrics, which 
are both type O.}
The Schwarzschild, Reissner–Nordstr\"om, and Kerr black hole solutions of the vacuum Einstein equations are all {algebraically special}, and of Petrov type D \cite{Kerr1963,Kinnersley1969}, while plane gravitational waves (in the far field zone of a gravitational wave emitting object) are of the algebraically special Petrov type N, with only $\Psi_4$ nonvanishing.

\section{Analogue gravity metric} 
The equation of motion of the linearized perturbation field in a barotropic, irrotational, and inviscid flow is analogous to a massless, minimally coupled scalar field in curved spacetime \cite{unruh, BLV}. The effective metric, the {\em 
physical acoustic metric}, in $3+1$D has the form\footnote{{We adopt the notational conventions of Ref.~\cite{Datta_2022},  
using a subscript 0 in round brackets for background variables such as $c_{s(0)}$ 
which are in general renormalized by the nonlinearity of fluid dynamics, 
while $c_{s0}$ denotes its strictly linearized 
counterpart.}}
\begin{equation}\label{agmn}
\mathfrak{g}_{\mu\nu}=\frac{\rho_{(0)}}{c_{s(0)}}\begin{bmatrix}
 -(c_{s(0)}^{2}-v_{(0)}^{2}) & \vdots & -v_{(0)}^{j} \\
\cdots&\cdots&\cdots\cdots \\
-v_{(0)}^{j}&\vdots &\delta_{ij}
\end{bmatrix},
\end{equation}
where  $\rho_{(0)}$ is the fluid density $\rho$ of the background, 
$c_{s(0)}^2=\frac{dp}{d\rho}|_{\rho=\rho_{(0)}}$ is the sound speed associated to this background,  
and $p=p(\rho)$ is the barotropic equation of state of the fluid. 
{We emphasize that all variables in \eqref{agmn} for this {\em physical} metric as well as in the metrics below 
depend on the three-dimensional position space coordinates and on time, unless otherwise specified.}

For a short wavelength perturbation, that is in the limit of geometrical acoustics, the conformal factor ${\rho_{(0)}}/{c_{s(0)}}$ is irrelevant because the phonon ray follows a null geodesic, $ds^2=0$. Therefore, the metric 
without the ${\rho_{(0)}}/{c_{s(0)}}$ factor determines the sound ray path in the fluid medium.
This has been coined the acoustic metric in the geometric limit \cite{BLV}. Using a tilde to denote
the geometric limit being taken by removing the conformal factor from \eqref{agmn},  
\begin{equation}
\tilde{\mathfrak g}_{\mu\nu}\coloneqq \begin{bmatrix}
 -(c_{s(0)}^{2}-v_{(0)}^{2}) & \vdots & -v_{(0)}^{j} \\
\cdots&\cdots&\cdots\cdots \\
-v_{(0)}^{j}&\vdots &\delta_{ij}
\end{bmatrix}.
\end{equation}
The inverse of the geometric acoustic metric reads 
\begin{equation}
\tilde{\mathfrak{g}}^{\mu\nu}
=\begin{bmatrix}
 -1 & \vdots & -v_{(0)}^{j} \\
\cdots&\cdots&\cdots\cdots \\
-v_{(0)}^{j}&\vdots &-(c_{s(0)}^{2}\delta_{ij}-v_{(0)}^{i}v_{(0)}^{j})
\end{bmatrix}.
\end{equation}
{Because the Petrov classification does not depend on a conformal factor in front of the metric, the 
geometric acoustic metric suffices to Petrov-classify the Weyl tensor of spacetime.}

\section{Flow Geometries with a global Petrov type}
\subsection{Type O spacetimes}
\subsubsection{Acoustic analogue of Minkowski spacetime}
Evidently, from Eq.~\eqref{gmnl}, a uniform static medium represents the acoustic analogue of Minkowski spacetime. 
Since the Weyl tensor identically vanishes, we have Petrov type O. 
\subsubsection{Acoustic analogue of Friedmann–Lema\^\i tre–Robertson–Walker metric}
An analogue of the isotropically expanding or contracting Universe described by the Friedmann–Lema\^\i tre–Robertson–Walker metric (up to a conformal factor), for a  Bose-Einstein condensate in a time-dependent spherically symmetric 
 trap, and a suitably time dependent contact interaction, has been derived in Refs.~\cite{BLV2003PRA,CPP}. 
The geometric acoustic metric line element 
can 
be written as 
\begin{equation}
ds^{2}=\tilde{\mathfrak g}_{\mu\nu}dx^\mu dx^\nu
=
{-c_{s(0)}^{2}dt^{2}+b^{2}d{r_{b}}^{2}+b^{2}{r_{b}}^{2}d\Omega^{2}}
.
\label{frg}
\end{equation}
where the sound speed $c_{s(0)}$ is a constant (near the center of the trap),  
and $r_b=r/b(t)$ is the scaled radial distance, with $b(t)$ defining the expansion of the superfluid gas.  
Using conformal time $\eta \coloneqq \int \frac{dt}{b(t)}$,  
this metric is manifestly conformally flat, and thus all Weyl tensor components are zero, just 
as for the Einstein-gravity cosmological counterpart.




\subsection{Type D: Irrotational vortex}
A vortex with constant circulation $\Gamma = \oint {\bm v}\cdot d{\bm s}$
around the vortex center has the azimuthal flow speed $\Gamma/(2\pi r)$.  
In the incompressible background approximation, 
[putting the (very large) 
sound speed equal to unity], the geometric acoustic metric then reads 
\begin{equation}
   \tilde{\mathfrak g}_{\mu\nu}= \begin{bmatrix}
        -1+\frac{\Gamma^2}{4\pi^2 r^2} & -\frac{\Gamma}{2\pi}&0&0\smallskip\\
        -\frac{\Gamma}{2\pi} & r^2 & 0 & 0 \smallskip\\
        0 & 0 & 1 & 0\smallskip\\
        0 & 0 & 0 & 1
    \end{bmatrix} .
\end{equation}
By assuming an incompressible background, we also neglect a possible density variation in the vortex core.

{The null tetrad vectors are here chosen to be} 
\begin{align}
    l_\mu &= \frac{1}{\sqrt{2}}(1-\Gamma/(2\pi r),r,0,0),\\
    n_\mu &= \frac{1}{\sqrt{2}}(1+\Gamma/(2\pi r),-r,0,0),\\
    m_\mu &= \frac{1}{\sqrt{2}}(0,0,1,i),\quad 
    \Bar{m}_\mu = \frac{1}{\sqrt{2}}(0,0,1,-i).
\end{align}   
The only nonvanishing Weyl scalars corresponding to this tetrad are
    \begin{align}
        \Psi_0 = -\frac{\Gamma^2}{2\pi ^2 r^4},\quad 
        \Psi_2 = \frac{\Gamma^2}{6\pi^2 r^4},\quad 
        \Psi_4 = -\frac{\Gamma^2}{2\pi^2 r^4}.
    \end{align}
For the vortex, the ${\mathcal Q}$ matrix then assumes the fully diagonal Jordan normal form 
\begin{equation}
    {\mathcal Q} = \begin{bmatrix}
        \frac{2\Gamma^2}{3\pi^2 r^4}& 0 & 0\\
        0 & -\frac{\Gamma^2}{3\pi^2r^4} & 0 \\
        0 & 0 & -\frac{\Gamma^2}{3\pi^2r^4}
    \end{bmatrix}.
\end{equation}
From Table \ref{Qtable}, we thus conclude the ${\mathcal Q}$ matrix again corresponds to Petrov type D.
{We note that here and in the spherically and cylindrically symmetric 
examples below, the curvature 
diverges at the origin within the incompressible background approximation.} 

\section{Analogue gravity spacetimes with no global Petrov type} 
\label{noglobal}
We consider below classes of 
background geometries which possess (in general) no 
global Petrov type.

\subsection{Background at rest} 
\subsubsection{Density variation along a specific direction}
A medium with a prescribed density variation along $x$ axis yields in $3+1$D 
the acoustic metric
\begin{equation}\label{gmns}
\mathfrak{g}_{\mu\nu}= \left(\frac{\rho_{(0)}(x,t)}{c_{s(0)}(x,t)} \right)\begin{bmatrix}
 -c^2_{s(0)}(x,t) & 0 & 0 & 0\\
0 & 1& 0 & 0 \\
0 & 0 & 1 & 0\\
0 & 0 &0 &1
\end{bmatrix}.
\end{equation}
The Weyl scalars are found by constructing the following 
tetrad vectors from the geometric acoustic metric, 
\begin{equation}\label{gmnl}
\tilde{\mathfrak g}_{\mu\nu}= \begin{bmatrix}
 -c^2_{s(0)}(x,t) & 0 & 0 & 0\\
0 & 1& 0 & 0 \\
0 & 0 & 1 & 0\\
0 & 0 &0 &1
\end{bmatrix}.
\end{equation}
Tetrad null vectors satisfying Eqs.~\eqref{ll}-\eqref{lm} are 
\begin{align}
& l_\mu= \frac{1}{\sqrt{2}}(c_{s(0)},1,0,0),\\
& n_\mu= \frac{1}{\sqrt{2}}(c_{s(0)},-1,0,0),\\
& m_\mu=  \frac{1}{\sqrt{2}}(0,0,1,{i}),\quad 
\bar{m}_{\mu}= \frac{1}{\sqrt{2}}(0,0,1,-{i}). 
\end{align}

We determine the Weyl scalars by employing the formula \eqref{Rt} for $D=4$, using Mathematica. 
The only nonvanishing Weyl scalar is given by 
\begin{equation}
\Psi _2=-\frac{\partial_x ^2 c_{s(0)}(x,t) }{6c_{s(0)}(x,t)}.
\end{equation} 
{Therefore the acoustic spacetime metric for a medium at rest, with a density variation 
along a specific direction, is Petrov type D. The possible exception is a linear variation of $c_{s(0)}$ 
with distance, then it is globally type O ($\Psi_2$ vanishes).  Finally, when the radial dependence of $c_{s(0)}$ has 
saddle point(s), this spacetime is locally type O.}  

{Since $\Psi_0$ is zero for the above choice of null tetrad vectors, 
$l^\mu$ and $n^\mu$ correspond to the two principal null directions.}

\subsubsection{Spherically symmetric density distribution}
The geometric acoustic metric in spherical polar coordinates $(r,\theta, \phi)$ is now given by
\begin{equation}\label{gmnsStatic}
\tilde{\mathfrak g}_{\mu\nu}= \begin{bmatrix}
 -c^2_{s(0)}(r,t) & 0 & 0 & 0\\
0 & 1& 0 & 0 \\
0 & 0 & r^2 & 0\\
0 & 0 &0 &r^2 \sin ^2\theta
\end{bmatrix}.
\end{equation}
{The tetrad null vectors satisfying Eqs.~\eqref{ll}-\eqref{lm} are now chosen to be} 
\begin{align}
& l_\mu= \frac{1}{\sqrt{2}}(c_{s(0)}(r,t),1,0,0),\\
& n_\mu= \frac{1}{\sqrt{2}}(c_{s(0)}(r,t),-1,0,0),\\
& m_\mu=  \frac{1}{\sqrt{2}}(0,0,r,{i}r\sin\theta),\quad 
 \bar{m}_{\mu}= \frac{1}{\sqrt{2}}(0,0,r,-{i}r\sin \theta).
\end{align}
The only nonvanishing Weyl scalar is 
\begin{equation}
\Psi _2=\frac{\partial _r c_{s(0)}(r,t)-r \partial _r ^2 c_{s(0)}(r,t)}{6r c_{s(0)}(r,t)}.
\end{equation}
Therefore this spacetime is again {of Petrov type D except when 
$\Psi_2$ vanishes, where it turns to type O. 
The latter happens at a given time whenever the spatial profile of $c_{s(0)}$  
is either locally or globally harmonic,  
 $c_{s(0)}= \frac12 C r^2 +D$,  
 where $C$ and $D$ are constants.  This profile can be obtained for example close to the center 
 of an isotropically harmonically trapped Bose-Einstein condensate, 
 for which the density is an inverted parabola within the Thomas-Fermi approximation.}

\subsection{One-dimensional flow}
The geometric metric associated with one-dimensional flow has the general form
\begin{equation}\label{mm}
\tilde{\mathfrak g}_{\mu\nu}(x,t)=  
\begin{bmatrix}
 & -(c_{s(0)}^{2}-v_{(0)}^{2}) & -v_{(0)} & 0 & 0\\
& -v_{(0)} & 1 & 0 & 0\\ 
& 0 & 0 & 1 &0 \\
 &0 & 0 & 0 &1\\
\end{bmatrix},
\end{equation}
where $c_{s(0)}=c_{s(0)}(x,t)$ and $v_{(0)}=v_{(0)}(x,t)$. 
{The null tetrad vectors, satisfying Eqs.~\eqref{ll}-\eqref{lm}, are here chosen to be} 
\begin{align}
& l_\mu= \frac{1}{\sqrt{2}}(c_{s(0)}-v_{(0)},1,0,0),\\
& n_\mu= \frac{1}{\sqrt{2}}(c_{s(0)}+v_{(0)},-1,0,0),\\
& m_\mu=  \frac{1}{\sqrt{2}}(0,0,1,{i}),\quad 
\bar{m}_{\mu}= \frac{1}{\sqrt{2}}(0,0,1,-{i}).
\end{align}


We consider as a specific example for a nonlinear plane wave travelling in a specific direction 
the so-called {\em simple} shock wave \cite{Landau1987Fluid}. 
The fluid is assumed to be equipped with a polytropic equation of state. 
$p(\rho) \propto \rho^\gamma$;  
for a Bose-Einstein condensate at zero temperature, $p= \frac12 g \rho^2$, with $g>0$ the two-body coupling strength.  
A simple wave propagating in the positive $x$ direction 
can be described by the Riemann wave equation 
 \cite{Riemann1860,Landau1987Fluid,Datta_2022} 
\begin{equation}\label{RW}
\frac{\partial v_{(0)}}{\partial t}+\left[c_{s0}+\left(\frac{\gamma +1}{2}\right) v_{(0)}\right]\frac{\partial v_{(0)}}{\partial x}=0,  
\end{equation}
where $c_{s0}$ is a positive number characterizing the sound speed for a linearized perturbation. 
Eq.~\eqref{RW} is derived from the irrotational inviscid fluid equations \cite{Landau1987Fluid}.
For a {\em simple} wave, the travelling wave can be described by a single quantity $v_{(0)}$, all 
other fluid variables are algebraic functions of $v_{(0)}(x,t)$.  
In particular, the fluid density is related to the velocity by
\begin{eqnarray}
\label{rhov}
\rho_{(0)}=\rho_{0}\left[1+\left(\frac{\gamma -1}{2}\right)\frac{v_{(0)}}{c_{s_{0}}}\right]^{\frac{2}{\gamma-1}}.
\end{eqnarray}
Therefore $\rho_{(0)}=\rho_0$ when  $v_{(0)}=0$. 
We also have a simple algebraic relationship for the sound speed,
\begin{equation}\label{cs}
c_{s(0)}=c_{s0}\left[1+\left(\frac{\gamma -1}{2}\right)\frac{v_{(0)}}{c_{s0}}\right].
\end{equation}
The solution of Eq.~\eqref{rhov} 
becomes multivalued when a discontinuity develops, resulting in a  shock wave 
solution \cite{Landau1987Fluid}. 

The resulting expression for $\Psi_2$ is fairly simple as well,
\begin{equation}
\Psi_2=-\frac{(1+\gamma)}{12c_{s_{0}}}\partial _x^2 v_{(0)}.
\end{equation}
{The spacetime is thus type D except at wavefront planes for which 
$\partial _x^2 v_{(0)}$ vanishes,  
 where it becomes type O.}


\subsection{Spherically symmetric background} \label{spherical} 

\begin{figure}[b]]\hspace*{-2em}
\centering
\includegraphics[width=\columnwidth]{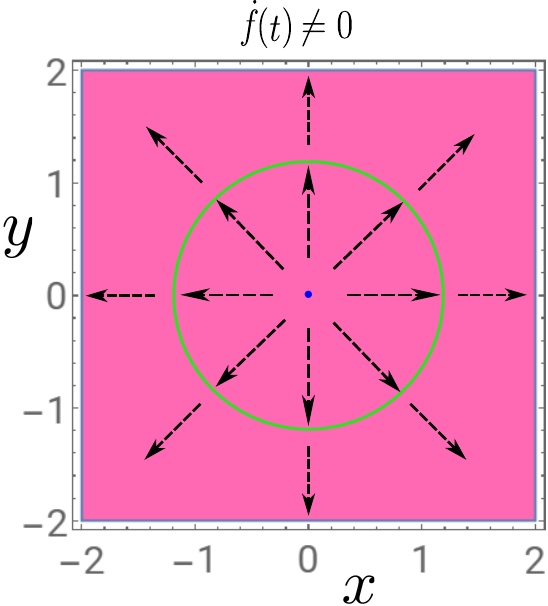}
\caption{{\em Petrov types for a spherically symmetric and time dependent background.}  
The dashed arrows represent the local direction of the (approximately incompressible) background flow. 
The prevailing Petrov type D spacetime is represented in pink. 
The green circle represents an isolated Petrov type O hypersurface. 
The scale for the $x$ and $y$ coordinates may be arbitrarily chosen.}
\label{FigSphere}
\end{figure} 

The geometric metric in spherical polar coordinates $(t,r,\theta,\phi)$ now is given by 
[here and below we again put the 
(very large) sound speed equal to unity], 
\begin{equation}\label{mmb}
\tilde{\mathfrak g}_{\mu\nu}(r,t)=  
\begin{bmatrix}
 & -(1-v_{(0)}^{2}) & -v_{(0)} & 0 & 0\\
& -v_{(0)} & 1 & 0 & 0\\ 
& 0 & 0 & r^2 &0 \\
 &0 & 0 & 0 & r^2\sin ^2 \theta \\
\end{bmatrix},
\end{equation}
where  $v_{(0)}=v_{(0)}(r,t)$. 
Because of continuity,  
\begin{equation}
v_{(0)}(r,t)=\frac{f(t)}{r^2}, 
\end{equation}
where $f(t)$ can be any function of time.

We now employ the following null tetrad vectors, as usual 
satisfying Eqs.~\eqref{ll}-\eqref{lm}, 
\begin{align}
& l_\mu= \frac{1}{\sqrt{2}}(1-v_{(0)},1,0,0),\\
& n_\mu= \frac{1}{\sqrt{2}}(1+v_{(0)},-1,0,0),\\
& m_\mu=  \frac{1}{\sqrt{2}}(0,0,1,{i}),\quad
\bar{m}_{\mu}= \frac{1}{\sqrt{2}}(0,0,1,-{i}).
\end{align}
All the Weyl scalars are zero except $\Psi_2$, 
\begin{equation}\label{psi2r}
\Psi_2=\frac1{2r^2} \left(5v_{(0)}^{2}-\frac{1}{r}\frac{df}{dt}\right).
\end{equation}
{\em Time independent flow.} 
From the above expression for $\Psi_2$, a time independent flow 
[$f(t)={\rm constant}$]  corresponds to a stationary metric, $\Psi_2=5v_{(0)}^{2}/{(2r^2)}$, 
and the metric is globally type D. 
 

{\em Time dependent flow.} 
It is possible in this case to engineer an isolated Petrov type O sphere at some radius $r_{\rm O}$, such that at $r=r_{\rm O}$, $\psi_2$ vanishes. From Eq.~\eqref{psi2r}, we have 
\begin{equation}
r_{\rm O}=\left(\frac{5f^2}{\dot{f}}\right)^\frac 13.
\end{equation}
Alternatively, fixing a radius $r_{\rm O}$, one determines a corresponding  $f(t)$ 
to produce an isolated Petrov type O spherical hypersurface. 
From Eq.~\eqref{psi2r}, we have the following first order differential equation for $f$:
\begin{equation} 
\frac{df}{dt}-\frac{5}{r_{\rm O}^3} f^2=0 ,
\end{equation}
which is readily solved by 
\begin{equation}
f=\left(-\frac{5t}{r_{\rm O}^3}+\frac{1}{f_0}\right)^{-1}, 
\end{equation}
where $f=f_0$ at $t=0$. We display the Petrov types in Fig.~\ref{FigSphere}. 

{The radially pulsating analogue spacetime will in general radiate the longitudinal acoustic 
analogue of spherical gravitational waves.} 


\subsection{Streaming motion past a cylinder}
\label{streaming}

\begin{figure}[t]\hspace*{-2em}
\centering
\includegraphics[width=\columnwidth]{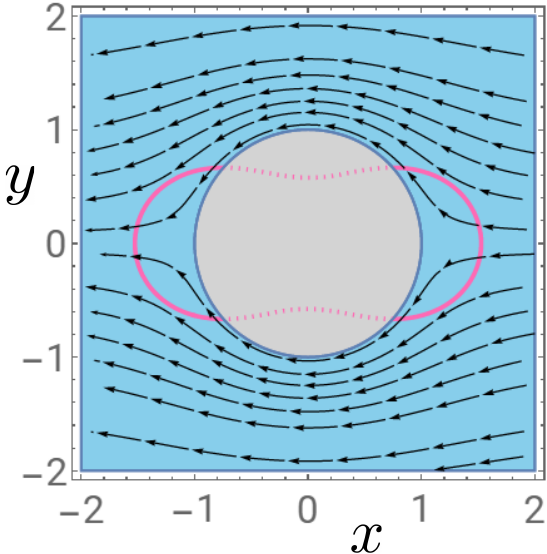} 
\caption{{\em Petrov types of streaming motion past a cylinder.} 
The gray area represents  the impenetrable 
cylinder. The pink line represents the isolated cylindrically symmetric Petrov type D hypersurface 
according to Eq.~\eqref{quadratic} (dashing
representing its unphysical part inside the cylinder), immersed in a Petrov type I  
spacetime represented by sky blue. The $x$ and $y$ coordinates are scaled by the cylinder radius $a$.} 
\label{FigCyl}
\end{figure}

We consider an incompressible background flow 
which has already been employed in the analogue gravity context by Ref.~\cite{FISCHER200322}.
The fluid here has velocity $U>0$ at infinite distance, from right to left, and is moving past an impenetrable 
cylinder of radius $a$. The maximal velocity on the cylinder surface (see Fig.~\ref{FigCyl}) is $2U$. 
For the incompressibility approximation to be applicable, one therefore needs $U\ll \frac12$.

Using the conformal transformation techniques familiar 
from hydrodynamics in two spatial dimensions, the two-dimensional velocity components are found to be 
\begin{align}
v_x (x,y) &= -U\left(1+\frac{a^2(y^2-x^2)}{r^4}\right),\\
v_y(x,y) &= 2U\frac{xya^2}{r^4}, 
\end{align}
where $r^2=x^2+y^2$. 
{We then choose the null tetrad vectors to be} 
\begin{align}
    l_\mu &= \frac{1}{\sqrt{2}}(1,0,0,1),\quad
        n_\mu = \frac{1}{\sqrt{2}}(1,0,0,-1),\\
    m_\mu &= \frac{1}{\sqrt{2}}(-v_x-i v_y,1,i,0),\\
    \Bar{m}_\mu &= \frac{1}{\sqrt{2}}(-v_x+i v_y,1,-i,0).
\end{align}    
For our choice of tetrad, we now have $\Psi_0\neq 0$, therefore we have to compute the 
complete set of eigenvalues of the ${\mathcal Q}$ matrix. 
The nonzero Weyl scalars are
\begin{align}
    \Psi_0&=-\frac{3 a^2 U^2 \left(a^2-(x+i y)^2\right)}{(x+i y)^2 (y+i x)^4},\\
    \Psi_2 &=\frac{4 a^4 U^2}{3r^6},\\
   \Psi_4 &=-\frac{3 a^2 U^2 \left(a^2-(x-i y)^2\right)}{(x-i y)^2 (x+i y)^4}.
\end{align}
As a result, the eigenvalues of the ${\mathcal Q}$ matrix are (using $x = r\cos\theta,\,y=r\sin\theta$), 
\begin{align}
    \lambda_1 &= -\frac{8a^4U^2}{3r^6},\\
    \lambda_2 &= \frac{a^2U^2 (4a^2-9\sqrt{a^4-2a^2r^2\cos(2\theta)+r^4})}{3r^6},\\
    \lambda_3 &= \frac{a^2U^2 (4a^2+9\sqrt{a^4-2a^2r^2\cos(2\theta)+r^4})}{3r^6}.
\end{align}

Searching for a possible degeneracy of the ${\mathcal Q}$ matrix eigenvalues, 
if $\lambda_2=\lambda_3$,  one would have $a^4-2a^2r^2\cos(2\theta)+r^4=0$.
But this is zero only for the two stagnation points 
$\theta=0,\pi$ at $r=a$, {i.e., on the surface of the
impenetrable cylinder}. 
Furthermore, since $\lambda_1<0$ and $\lambda_3>0$, they cannot possibly be equal.
Turning to $\lambda_1=\lambda_2$, we obtain 
$4a^2 = 3\sqrt{a^4-2a^2r^2\cos(2\theta)+r^4}$,
or equivalently a quadratic equation for $r^2$, 
\begin{equation}
    9r^4-18a^2r^2\cos(2\theta)-7a^4=0. \label{quadratic} 
\end{equation}
{The corresponding curve $r(\theta)$ yields an isolated algebraically special 
hypersurface of Petrov type D, see the peanut-shaped pink curve in Fig.~\ref{FigCyl}, surrounded by the 
algebraically general Petrov type I acoustic spacetime of the flow past the cylinder,  
which covers the remaining fluid.} 
{We note that for any cylinder radius $a$ the shape {and relative size}  
of the pink curve in  Fig.~\ref{FigCyl} remains invariant, and that the value of any nonzero $U$ (within the confines of the incompressibility approximation) does not have an effect on the Petrov types of this flow.}

{We finally remark that for the cylindrically symmetric version of the spherically symmetric incompressible 
background fluid
of section \ref{spherical}, and thus now with $v_{(0)}(r,t)=f(t)/r$, we find for $\dot f(t)\neq 0$ that a 
predominantly type I spacetime is obtained (instead of type D for the spherically symmetric case), as for the present streaming motion past a cylinder, and also the isolated hypersurface is again type D (instead of type O).}

\section{Petrov types of Painlev\'e-Gullstrand geometries}

{One may conjecture that increasingly complex and possibly also time-dependent flows potentially 
produce other Petrov types, different from those obtained in the above \footnote{We note in this regard that the primary constraint on the 
possible Petrov types of a general spatially three-dimensional flow 
is that constant time slices of the analogue metric must be conformally flat in the Newtonian lab,
as follows from their Painlev\'e-Gullstrand form,  
in distinction to a general solution of the Einstein equations, where no such condition applies.}.
 This is however in general not the case: We provide in what follows the proof that 
{for all spacetimes constructed from an incompressible background, and for all quasi-one-dimensional 
 compressible flows
 the Petrov types O, D, and I describe all corresponding Painlev\'e-Gullstrand metric analogue spacetimes.} 
 The proof is based on the fact that if the $\mathcal Q$ matrix is real 
 in addition to being symmetric, it can always be diagonalized, and due to its also being traceless the preceding 
 statement on the possible Petrov types follows.}

The quasiparticles in the analogue spacetime experience an effective geometry with a metric of the 
Painlev\'e-Gullstrand form 
\begin{equation}\label{PGmetric}
    {\rm d}s^2 = \frac{\rho}{c_{s(0)}}\left[-c_{s(0)}^2{\rm d}t^2+\delta_{ij}({\rm d}x^i-v^i{\rm d}t)({\rm d}x^j-v^j{\rm d}t)\right].
\end{equation}
%
For computational convenience, we will omit the conformal factor $\rho/c_{s(0)}$ in the following. 

{We will work in this section in an orthonormal frame. 
A natural choice for the orthonormal one-form corresponding to \eqref{PGmetric} 
is given by} 
\begin{equation}
    e^{\hat{0}} = \mathrm{d}t, \quad e^{\hat{i}} = \mathrm{d}x^i -v^i\mathrm{d}t,
\end{equation}
where $i\in\{1,2,3\}$. 
Based on this choice, we will now determine the conditions for the $\mathcal{Q}$ matrix to be real based on the corresponding orthonormal tetrad components of the Weyl and Riemann tensors.
{We note here that the complex null tetrad of section \ref{nullandWeyl} 
can be decomposed into the orthonormal tetrad basis as follows} 
\begin{align}\label{eq:nullortho}
    l &= \frac{1}{\sqrt{2}}\left(e^{\hat{t}}+e^{\hat{1}}\right), &\quad
    n = \frac{1}{\sqrt{2}}\left(e^{\hat{t}}-e^{\hat{1}}\right),\\
    m &= \frac{1}{\sqrt{2}}\left(e^{\hat{2}}+ie^{\hat{3}}\right), & \quad
    \Bar{m} = \frac{1}{\sqrt{2}}\left(e^{\hat{2}}-ie^{\hat{3}}\right).
\end{align}

The $\mathcal Q$ matrix \eqref{Qmatrix} being real relies on the following three conditions,
\begin{itemize}
    \item[$\diamond$] $\Psi_2\in\mathbb{R}$,
    \item[$\diamond$] $\Psi_0=\Bar{\Psi}_4$,
    \item[$\diamond$] $\Psi_1=-\Bar{\Psi}_3$.
\end{itemize}
{It should be noted in this connection that the coordinate basis components of the Weyl tensor 
are real numbers.}
By substituting Eq.~\eqref{eq:nullortho} into each of the above conditions, 
simple algebraic manipulations yield  
\begin{align}
    C_{\lambda\mu\nu\kappa}l^{\lambda}n^{\mu}e_{\hat{2}}^{\nu}e_{\hat{3}}^{\kappa}&=0,\\
    C_{\lambda\mu\nu\kappa}e_{\hat{t}}^{\lambda}m^{\mu}e_{\hat{1}}^{\nu}m^{\kappa}&=0,\\
    C_{\lambda\mu\nu\kappa}l^{\lambda}n^{\mu}e_{\hat{1}}^{\nu}m^{\kappa} &= 0.
\end{align}
In addition, utilizing the symmetries of the Weyl tensor and Eq.\eqref{eq:nullortho}, we obtain
\begin{align}
    C_{\hat{0}\hat{1}\hat{2}\hat{3}} &= 0,\\
    C_{\hat{0}\hat{2}\hat{1}\hat{2}}-C_{\hat{0}\hat{3}\hat{1}\hat{3}}
    +i(C_{\hat{0}\hat{2}\hat{1}\hat{3}}+C_{\hat{0}\hat{3}\hat{1}\hat{2}}) &=0,\\
    C_{\hat{0}\hat{1}\hat{1}\hat{2}}+iC_{\hat{0}\hat{1}\hat{1}\hat{3}} &= 0,
\end{align}
where $C_{\hat{a}\hat{b}\hat{c}\hat{d}}\equiv C_{\lambda\mu\nu\kappa}e_{\hat{a}}^{\lambda}e_{\hat{b}}^{\mu}e_{\hat{c}}^{\nu}e_{\hat{d}}^{\kappa}$. Since the $C_{\hat{a}\hat{b}\hat{c}\hat{d}}$ are real, in total five conditions imposed on the Weyl tensor components arise. 
Furthermore, each of these conditions can be directly expressed in terms of the Riemann and Ricci tensor components in the same orthonormal tetrad. A summary of these conditions is provided in Table \ref{table:Condition}.

{\def\arraystretch{1.5}
\begin{table}[t]
\centering
\begin{tabular}{|c|c|c|}
\hline
Reality Condition & Weyl & Riemann  \\ \hline
$\Psi_2\in\mathbb{R}$ & $ C_{\hat{0}\hat{1}\hat{2}\hat{3}} = 0$ & $R_{\hat{0}\hat{1}\hat{2}\hat{3}}=0$\\ \hline
\multirow{ 2}{*}{$\Psi_0=\Bar{\Psi}_4$} & $ C_{\hat{0}\hat{2}\hat{1}\hat{2}}-C_{\hat{0}\hat{3}\hat{1}\hat{3}}=0$ & $R_{\hat{0}\hat{2}\hat{1}\hat{2}}-R_{\hat{0}\hat{3}\hat{1}\hat{3}}=0$ \\
 & $C_{\hat{0}\hat{2}\hat{1}\hat{3}}+C_{\hat{0}\hat{3}\hat{1}\hat{2}}=0$ & $R_{\hat{0}\hat{2}\hat{1}\hat{3}}+R_{\hat{0}\hat{3}\hat{1}\hat{2}}=0$ \\ \hline
 \multirow{ 2}{*}{$\Psi_1=-\Bar{\Psi}_3$} & $C_{\hat{0}\hat{1}\hat{1}\hat{2}}=0$ & $R_{\hat{0}\hat{1}\hat{1}\hat{2}}+R_{\hat{0}\hat{2}}/2=0$ \\
 & $C_{\hat{0}\hat{1}\hat{1}\hat{3}}=0$ & $R_{\hat{0}\hat{1}\hat{1}\hat{3}}+R_{\hat{0}\hat{3}}/2=0$ \\ \hline
\end{tabular}
\caption{Conditions for the \texorpdfstring{$\mathcal Q$}{q} matrix to be real in the orthonormal tetrad frame.}
\label{table:Condition}
\end{table}
}
For the conditions specified in Table~\ref{table:Condition}, only $R_{\hat{0}\hat{i}\hat{j}\hat{k}}$ and $R_{\hat{0}\hat{i}}$, where $i,j,k\in{1,2,3}$, are required. Furthermore, according to \cite{FISCHER200322}, for an iso-tachic speed of sound $c$  (independent of position and time), we have 
\begin{align}
    R_{\hat{0}\hat{i}\hat{j}\hat{k}} &= \partial_i\Omega_{jk},\\
    R_{\hat{0}\hat{i}} &= \partial_k\Omega_{ki},
\end{align}
where
\begin{equation}
    \Omega_{ij} = \frac{1}{2c_{s(0)}}(\partial_iv_j-\partial_jv_i),
\end{equation}
and $i,j\in{x,y,z}$ represent the coordinate basis indices. Given that $\Omega_{ij}=0$ for {irrotational flow (as required 
for the derivation of the minimally coupled scalar wave equation associated to $\mathfrak{g}_{\mu\nu}$ 
from perfect fluid dynamics),} the $\mathcal{Q}$ matrix is always real and hence diagonalizable: For an incompressible background flow,  the only possible Petrov types are thus type I, type O, and type D.

For a compressible fluid, one has 
\begin{align}
    R_{\hat{0}\hat{i}\hat{j}\hat{k}} = -\partial_jD_{ik}+\partial_kD_{ij},\\
    R_{\hat{0}\hat{i}} = -R_{\hat{0}\hat{k}\hat{k}\hat{i}} = -\partial_kD_{ki}+\partial_iD_{kk},
\end{align}
where 
\begin{equation}
    D_{ij} = \frac{1}{2c_{s(0)}}(\partial_iv_j+\partial_jv_i).
\end{equation}
{with $c_{s(0)}$ now in general position (and time) dependent.}
For {irrotational} flow, the Riemann and Ricci tensor components in the orthonormal tetrad basis become 
\begin{align}
    R_{\hat{0}\hat{i}\hat{j}\hat{k}} = (\partial_j \ln{c_{s(0)}})\,D_{ik}- (\partial_k\ln{c_{s(0)}})\,D_{ij},\\
    R_{\hat{0}\hat{i}} = -R_{\hat{0}\hat{k}\hat{k}\hat{i}} = (\partial_k \ln{c_{s(0)}})\,D_{ki}-(\partial_i\ln{c_{s(0)}})\,D_{kk}.
\end{align}
{The conditions obtained from Table  \ref{table:Condition} 
then do not hold in general for compressible backgrounds. However, for a (quasi-)one-dimensional system, most commonly studied in present analogue gravity experiments as well as in theory, one readily concludes that $ R_{\hat{0}\hat{i}\hat{j}\hat{k}}=R_{\hat{0}\hat{i}} = 0$, so that we are again led to the  
Petrov types I, O, and D.}

\section{Conclusion}

In our Petrov classification analysis of analogue spacetimes, which is 
based on the Weyl scalars corresponding to a null tetrad, 
we have found Petrov type O, type D, and type I spacetimes.
The most nontrivial, that is algebraically general, Petrov type I was indeed found for a background flow 
displaying the least flow symmetry, that is, streaming motion past an impenetrable cylinder.

 {Regarding black (dumb) hole geometries, when it comes to the acoustic Schwarzschild geometry, in 
spherical laboratory symmetry, it can be shown that it can be reproduced by fine-tuning the equation of state and the external forcing of the system, either without flow (Appendix of Ref.~\cite{Visser_2005}),  or with a 
spherically symmetric flow $v(r)\propto r^{-1/2}$ \cite{MattCQG}. 
The equatorial constant-time slice of the Kerr geometry, due to its asymptotic conformal flatness,  
 and again under very specifically engineered conditions, can be mimicked 
 by a (compressible) vortex flow \cite{Visser_2005}. 
Both of these acoustic geometries are then Petrov type D, 
as their Einstein-gravitational counterparts are.} 

{The Lense-Thirring spacetime, obtained as an approximate solution of the Einstein vacuum equations outside the surface of a rotating star \cite{Lense},  can be written asymptotically, 
with moderate modifications, in the Painlev\'e-Gullstrand form
 \cite{Baines}. This asymptotic 
 variant of the Lense-Thirring 
metric has been demonstrated to be Petrov type I, now with isolated hypersurfaces at pole and equator, respectively \cite{Baines}.
It thus falls under the same algebraically general class as the streaming motion past a cylinder we discussed in section \ref{streaming}.  
Given that, again asymptotically, 
the Kerr spacetime merges into the Lense-Thirring spacetime, this further illustrates the sensitivity of the Petrov classification even to relatively small deviations from a given metric structure.}


We note that the Petrov classification applies to any spacetime in general. 
However, since in nonrelativistic analogue setups, the metric does not represent a solution of the 
Einstein equations, further subclasses are conceivable by employing also the Ricci part of the Riemann curvature tensor in  Eq.~\eqref{Rt}, cf.~Refs.~\cite{Plebanski,Hall,McIntosh}. 

Finally, also considering, then, a general-relativistic curved spacetime background given by a solution of the Einstein equations 
(in vacuum or with matter), we will have two spacetime metrics, one corresponding to the real spacetime and the other  
to the acoustic metric for linearized perturbations cf., e.g., Refs.~\cite{Moncrief,Neven,Molina-Paris,Nakahara}. 
Such a general relativistic background can then reveal a potential interplay of 
two Petrov types, those for real and for analogue spacetimes, respectively.
\pagebreak

\section{Acknowledgments} 
This work has been supported by the National Research Foundation of Korea under 
Grants No.~2017R1A2A2A05001422 and No.~2020R1A2C2008103.

\bibliography{petrov12}

\begin{thebibliography}{53}%
\makeatletter
\providecommand \@ifxundefined [1]{%
 \@ifx{#1\undefined}
}%
\providecommand \@ifnum [1]{%
 \ifnum #1\expandafter \@firstoftwo
 \else \expandafter \@secondoftwo
 \fi
}%
\providecommand \@ifx [1]{%
 \ifx #1\expandafter \@firstoftwo
 \else \expandafter \@secondoftwo
 \fi
}%
\providecommand \natexlab [1]{#1}%
\providecommand \enquote  [1]{``#1''}%
\providecommand \bibnamefont  [1]{#1}%
\providecommand \bibfnamefont [1]{#1}%
\providecommand \citenamefont [1]{#1}%
\providecommand \href@noop [0]{\@secondoftwo}%
\providecommand \href [0]{\begingroup \@sanitize@url \@href}%
\providecommand \@href[1]{\@@startlink{#1}\@@href}%
\providecommand \@@href[1]{\endgroup#1\@@endlink}%
\providecommand \@sanitize@url [0]{\catcode `\\12\catcode `\$12\catcode
  `\&12\catcode `\#12\catcode `\^12\catcode `\_12\catcode `\%12\relax}%
\providecommand \@@startlink[1]{}%
\providecommand \@@endlink[0]{}%
\providecommand \url  [0]{\begingroup\@sanitize@url \@url }%
\providecommand \@url [1]{\endgroup\@href {#1}{\urlprefix }}%
\providecommand \urlprefix  [0]{URL }%
\providecommand \Eprint [0]{\href }%
\providecommand \doibase [0]{https://doi.org/}%
\providecommand \selectlanguage [0]{\@gobble}%
\providecommand \bibinfo  [0]{\@secondoftwo}%
\providecommand \bibfield  [0]{\@secondoftwo}%
\providecommand \translation [1]{[#1]}%
\providecommand \BibitemOpen [0]{}%
\providecommand \bibitemStop [0]{}%
\providecommand \bibitemNoStop [0]{.\EOS\space}%
\providecommand \EOS [0]{\spacefactor3000\relax}%
\providecommand \BibitemShut  [1]{\csname bibitem#1\endcsname}%
\let\auto@bib@innerbib\@empty
\bibitem [{\citenamefont {Unruh}(1981)}]{unruh}%
  \BibitemOpen
  \bibfield  {author} {\bibinfo {author} {\bibfnamefont {W.~G.}\ \bibnamefont
  {Unruh}},\ }\bibfield  {title} {\bibinfo {title} {{Experimental Black-Hole
  Evaporation?}},\ }\href {https://doi.org/10.1103/PhysRevLett.46.1351}
  {\bibfield  {journal} {\bibinfo  {journal} {Phys. Rev. Lett.}\ }\textbf
  {\bibinfo {volume} {46}},\ \bibinfo {pages} {1351} (\bibinfo {year}
  {1981})}\BibitemShut {NoStop}%
\bibitem [{\citenamefont {Trautman}(1966)}]{Trautman}%
  \BibitemOpen
  \bibfield  {author} {\bibinfo {author} {\bibfnamefont {A.}~\bibnamefont
  {Trautman}},\ }\bibinfo {title} {{{Comparison of Newtonian and Relativistic
  Theories of Space-time}}},\ in\ \href@noop {} {\emph {\bibinfo {booktitle}
  {{Perspectives in Geometry and Relativity', Essays in Honor of V\'aclav
  Hlavat\'y}}}}\ (\bibinfo  {publisher} {{Indiana University Press}},\ \bibinfo
  {year} {1966})\ pp.\ \bibinfo {pages} {413--425}\BibitemShut {NoStop}%
\bibitem [{\citenamefont {Barcel{\'o}}\ \emph {et~al.}(2011)\citenamefont
  {Barcel{\'o}}, \citenamefont {Liberati},\ and\ \citenamefont {Visser}}]{BLV}%
  \BibitemOpen
  \bibfield  {author} {\bibinfo {author} {\bibfnamefont {C.}~\bibnamefont
  {Barcel{\'o}}}, \bibinfo {author} {\bibfnamefont {S.}~\bibnamefont
  {Liberati}},\ and\ \bibinfo {author} {\bibfnamefont {M.}~\bibnamefont
  {Visser}},\ }\bibfield  {title} {\bibinfo {title} {{Analogue Gravity}},\
  }\href {https://doi.org/10.12942/lrr-2011-3} {\bibfield  {journal} {\bibinfo
  {journal} {Living Reviews in Relativity}\ }\textbf {\bibinfo {volume} {14}},\
  \bibinfo {pages} {3} (\bibinfo {year} {2011})}\BibitemShut {NoStop}%
\bibitem [{\citenamefont {Landau}\ and\ \citenamefont
  {Lifshitz}(1987)}]{Landau1987Fluid}%
  \BibitemOpen
  \bibfield  {author} {\bibinfo {author} {\bibfnamefont {L.~D.}\ \bibnamefont
  {Landau}}\ and\ \bibinfo {author} {\bibfnamefont {E.~M.}\ \bibnamefont
  {Lifshitz}},\ }\href {http://www.worldcat.org/isbn/0750627670} {\emph
  {\bibinfo {title} {Fluid Mechanics, Second Edition: Volume 6 (Course of
  Theoretical Physics)}}},\ \bibinfo {edition} {2nd}\ ed.\ (\bibinfo
  {publisher} {Butterworth-Heinemann},\ \bibinfo {year} {1987})\BibitemShut
  {NoStop}%
\bibitem [{\citenamefont {Lahav}\ \emph {et~al.}(2010)\citenamefont {Lahav},
  \citenamefont {Itah}, \citenamefont {Blumkin}, \citenamefont {Gordon},
  \citenamefont {Rinott}, \citenamefont {Zayats},\ and\ \citenamefont
  {Steinhauer}}]{PhysRevLett.105.240401}%
  \BibitemOpen
  \bibfield  {author} {\bibinfo {author} {\bibfnamefont {O.}~\bibnamefont
  {Lahav}}, \bibinfo {author} {\bibfnamefont {A.}~\bibnamefont {Itah}},
  \bibinfo {author} {\bibfnamefont {A.}~\bibnamefont {Blumkin}}, \bibinfo
  {author} {\bibfnamefont {C.}~\bibnamefont {Gordon}}, \bibinfo {author}
  {\bibfnamefont {S.}~\bibnamefont {Rinott}}, \bibinfo {author} {\bibfnamefont
  {A.}~\bibnamefont {Zayats}},\ and\ \bibinfo {author} {\bibfnamefont
  {J.}~\bibnamefont {Steinhauer}},\ }\bibfield  {title} {\bibinfo {title}
  {{Realization of a Sonic Black Hole Analog in a Bose-Einstein Condensate}},\
  }\href {https://doi.org/10.1103/PhysRevLett.105.240401} {\bibfield  {journal}
  {\bibinfo  {journal} {Phys. Rev. Lett.}\ }\textbf {\bibinfo {volume} {105}},\
  \bibinfo {pages} {240401} (\bibinfo {year} {2010})}\BibitemShut {NoStop}%
\bibitem [{\citenamefont {Mu{\~n}oz~de Nova}\ \emph {et~al.}(2019)\citenamefont
  {Mu{\~n}oz~de Nova}, \citenamefont {Golubkov}, \citenamefont {Kolobov},\ and\
  \citenamefont {Steinhauer}}]{Munoz}%
  \BibitemOpen
  \bibfield  {author} {\bibinfo {author} {\bibfnamefont {J.~R.}\ \bibnamefont
  {Mu{\~n}oz~de Nova}}, \bibinfo {author} {\bibfnamefont {K.}~\bibnamefont
  {Golubkov}}, \bibinfo {author} {\bibfnamefont {V.~I.}\ \bibnamefont
  {Kolobov}},\ and\ \bibinfo {author} {\bibfnamefont {J.}~\bibnamefont
  {Steinhauer}},\ }\bibfield  {title} {\bibinfo {title} {{Observation of
  thermal Hawking radiation and its temperature in an analogue black hole}},\
  }\href {https://doi.org/10.1038/s41586-019-1241-0} {\bibfield  {journal}
  {\bibinfo  {journal} {Nature}\ }\textbf {\bibinfo {volume} {569}},\ \bibinfo
  {pages} {688} (\bibinfo {year} {2019})}\BibitemShut {NoStop}%
\bibitem [{\citenamefont {Kolobov}\ \emph {et~al.}(2021)\citenamefont
  {Kolobov}, \citenamefont {Golubkov}, \citenamefont {Mu{\~n}oz~de Nova},\ and\
  \citenamefont {Steinhauer}}]{Steinhauer2021}%
  \BibitemOpen
  \bibfield  {author} {\bibinfo {author} {\bibfnamefont {V.~I.}\ \bibnamefont
  {Kolobov}}, \bibinfo {author} {\bibfnamefont {K.}~\bibnamefont {Golubkov}},
  \bibinfo {author} {\bibfnamefont {J.~R.}\ \bibnamefont {Mu{\~n}oz~de Nova}},\
  and\ \bibinfo {author} {\bibfnamefont {J.}~\bibnamefont {Steinhauer}},\
  }\bibfield  {title} {\bibinfo {title} {{Observation of stationary spontaneous
  Hawking radiation and the time evolution of an analogue black hole}},\ }\href
  {https://doi.org/10.1038/s41567-020-01076-0} {\bibfield  {journal} {\bibinfo
  {journal} {Nature Physics}\ }\textbf {\bibinfo {volume} {17}},\ \bibinfo
  {pages} {362} (\bibinfo {year} {2021})}\BibitemShut {NoStop}%
\bibitem [{\citenamefont {Volovik}(2009)}]{Grisha}%
  \BibitemOpen
  \bibfield  {author} {\bibinfo {author} {\bibfnamefont {G.~E.}\ \bibnamefont
  {Volovik}},\ }\href
  {https://doi.org/10.1093/acprof:oso/9780199564842.001.0001} {\emph {\bibinfo
  {title} {{{The Universe in a Helium Droplet}}}}}\ (\bibinfo  {publisher}
  {Oxford University Press},\ \bibinfo {year} {2009})\BibitemShut {NoStop}%
\bibitem [{\citenamefont {Fedichev}\ and\ \citenamefont
  {Fischer}(2003)}]{Fedichev}%
  \BibitemOpen
  \bibfield  {author} {\bibinfo {author} {\bibfnamefont {P.~O.}\ \bibnamefont
  {Fedichev}}\ and\ \bibinfo {author} {\bibfnamefont {U.~R.}\ \bibnamefont
  {Fischer}},\ }\bibfield  {title} {\bibinfo {title} {{Gibbons-Hawking Effect
  in the Sonic de Sitter Space-Time of an Expanding Bose-Einstein-Condensed
  Gas}},\ }\href {https://doi.org/10.1103/PhysRevLett.91.240407} {\bibfield
  {journal} {\bibinfo  {journal} {Phys. Rev. Lett.}\ }\textbf {\bibinfo
  {volume} {91}},\ \bibinfo {pages} {240407} (\bibinfo {year}
  {2003})}\BibitemShut {NoStop}%
\bibitem [{\citenamefont {Fischer}\ and\ \citenamefont
  {Sch\"utzhold}(2004)}]{Schuetzhold}%
  \BibitemOpen
  \bibfield  {author} {\bibinfo {author} {\bibfnamefont {U.~R.}\ \bibnamefont
  {Fischer}}\ and\ \bibinfo {author} {\bibfnamefont {R.}~\bibnamefont
  {Sch\"utzhold}},\ }\bibfield  {title} {\bibinfo {title} {{Quantum simulation
  of cosmic inflation in two-component Bose-Einstein condensates}},\ }\href
  {https://doi.org/10.1103/PhysRevA.70.063615} {\bibfield  {journal} {\bibinfo
  {journal} {Phys. Rev. A}\ }\textbf {\bibinfo {volume} {70}},\ \bibinfo
  {pages} {063615} (\bibinfo {year} {2004})}\BibitemShut {NoStop}%
\bibitem [{\citenamefont {Unruh}\ and\ \citenamefont
  {Sch{\"u}tzhold}(2007)}]{unruh2007quantum}%
  \BibitemOpen
  \bibfield  {author} {\bibinfo {author} {\bibfnamefont {W.~G.}\ \bibnamefont
  {Unruh}}\ and\ \bibinfo {author} {\bibfnamefont {R.}~\bibnamefont
  {Sch{\"u}tzhold}},\ }\href {https://doi.org/10.1007/3-540-70859-6} {\emph
  {\bibinfo {title} {{Quantum Analogues: From Phase Transitions to Black Holes
  and Cosmology}}}},\ \bibinfo {series} {Lecture Notes in Physics}, Vol.\
  \bibinfo {volume} {718}\ (\bibinfo  {publisher} {Springer Berlin
  Heidelberg},\ \bibinfo {year} {2007})\BibitemShut {NoStop}%
\bibitem [{\citenamefont {Ch\"a}\ and\ \citenamefont
  {Fischer}(2017)}]{PhysRevLett.118.130404}%
  \BibitemOpen
  \bibfield  {author} {\bibinfo {author} {\bibfnamefont {S.-Y.}\ \bibnamefont
  {Ch\"a}}\ and\ \bibinfo {author} {\bibfnamefont {U.~R.}\ \bibnamefont
  {Fischer}},\ }\bibfield  {title} {\bibinfo {title} {{Probing the Scale
  Invariance of the Inflationary Power Spectrum in Expanding
  Quasi-Two-Dimensional Dipolar Condensates}},\ }\href
  {https://doi.org/10.1103/PhysRevLett.118.130404} {\bibfield  {journal}
  {\bibinfo  {journal} {Phys. Rev. Lett.}\ }\textbf {\bibinfo {volume} {118}},\
  \bibinfo {pages} {130404} (\bibinfo {year} {2017})}\BibitemShut {NoStop}%
\bibitem [{\citenamefont {Hung}\ \emph {et~al.}(2013)\citenamefont {Hung},
  \citenamefont {Gurarie},\ and\ \citenamefont {Chin}}]{Chin}%
  \BibitemOpen
  \bibfield  {author} {\bibinfo {author} {\bibfnamefont {C.-L.}\ \bibnamefont
  {Hung}}, \bibinfo {author} {\bibfnamefont {V.}~\bibnamefont {Gurarie}},\ and\
  \bibinfo {author} {\bibfnamefont {C.}~\bibnamefont {Chin}},\ }\bibfield
  {title} {\bibinfo {title} {{From Cosmology to Cold Atoms: Observation of
  Sakharov Oscillations in a Quenched Atomic Superfluid}},\ }\href
  {https://doi.org/10.1126/science.1237557} {\bibfield  {journal} {\bibinfo
  {journal} {Science}\ }\textbf {\bibinfo {volume} {341}},\ \bibinfo {pages}
  {1213} (\bibinfo {year} {2013})}\BibitemShut {NoStop}%
\bibitem [{\citenamefont {Eckel}\ \emph {et~al.}(2018)\citenamefont {Eckel},
  \citenamefont {Kumar}, \citenamefont {Jacobson}, \citenamefont {Spielman},\
  and\ \citenamefont {Campbell}}]{Eckel}%
  \BibitemOpen
  \bibfield  {author} {\bibinfo {author} {\bibfnamefont {S.}~\bibnamefont
  {Eckel}}, \bibinfo {author} {\bibfnamefont {A.}~\bibnamefont {Kumar}},
  \bibinfo {author} {\bibfnamefont {T.}~\bibnamefont {Jacobson}}, \bibinfo
  {author} {\bibfnamefont {I.~B.}\ \bibnamefont {Spielman}},\ and\ \bibinfo
  {author} {\bibfnamefont {G.~K.}\ \bibnamefont {Campbell}},\ }\bibfield
  {title} {\bibinfo {title} {{A Rapidly Expanding Bose-Einstein Condensate: An
  Expanding Universe in the Lab}},\ }\href
  {https://doi.org/10.1103/PhysRevX.8.021021} {\bibfield  {journal} {\bibinfo
  {journal} {Phys. Rev. X}\ }\textbf {\bibinfo {volume} {8}},\ \bibinfo {pages}
  {021021} (\bibinfo {year} {2018})}\BibitemShut {NoStop}%
\bibitem [{\citenamefont {Viermann}\ \emph {et~al.}(2022)\citenamefont
  {Viermann}, \citenamefont {Sparn}, \citenamefont {Liebster}, \citenamefont
  {Hans}, \citenamefont {Kath}, \citenamefont {Parra-L{\'o}pez}, \citenamefont
  {Tolosa-Sime{\'o}n}, \citenamefont {S{\'a}nchez-Kuntz}, \citenamefont {Haas},
  \citenamefont {Strobel}, \citenamefont {Floerchinger},\ and\ \citenamefont
  {Oberthaler}}]{Viermann}%
  \BibitemOpen
  \bibfield  {author} {\bibinfo {author} {\bibfnamefont {C.}~\bibnamefont
  {Viermann}}, \bibinfo {author} {\bibfnamefont {M.}~\bibnamefont {Sparn}},
  \bibinfo {author} {\bibfnamefont {N.}~\bibnamefont {Liebster}}, \bibinfo
  {author} {\bibfnamefont {M.}~\bibnamefont {Hans}}, \bibinfo {author}
  {\bibfnamefont {E.}~\bibnamefont {Kath}}, \bibinfo {author} {\bibfnamefont
  {{\'A}.}~\bibnamefont {Parra-L{\'o}pez}}, \bibinfo {author} {\bibfnamefont
  {M.}~\bibnamefont {Tolosa-Sime{\'o}n}}, \bibinfo {author} {\bibfnamefont
  {N.}~\bibnamefont {S{\'a}nchez-Kuntz}}, \bibinfo {author} {\bibfnamefont
  {T.}~\bibnamefont {Haas}}, \bibinfo {author} {\bibfnamefont {H.}~\bibnamefont
  {Strobel}}, \bibinfo {author} {\bibfnamefont {S.}~\bibnamefont
  {Floerchinger}},\ and\ \bibinfo {author} {\bibfnamefont {M.~K.}\ \bibnamefont
  {Oberthaler}},\ }\bibfield  {title} {\bibinfo {title} {Quantum field
  simulator for dynamics in curved spacetime},\ }\href
  {https://doi.org/10.1038/s41586-022-05313-9} {\bibfield  {journal} {\bibinfo
  {journal} {Nature}\ }\textbf {\bibinfo {volume} {611}},\ \bibinfo {pages}
  {260} (\bibinfo {year} {2022})}\BibitemShut {NoStop}%
\bibitem [{\citenamefont {Weyl}(1918)}]{Weyl}%
  \BibitemOpen
  \bibfield  {author} {\bibinfo {author} {\bibfnamefont {H.}~\bibnamefont
  {Weyl}},\ }\bibfield  {title} {\bibinfo {title} {{Reine
  Infinitesimalgeometrie}},\ }\href {https://doi.org/10.1007/BF01199420}
  {\bibfield  {journal} {\bibinfo  {journal} {Mathematische Zeitschrift}\
  }\textbf {\bibinfo {volume} {2}},\ \bibinfo {pages} {384} (\bibinfo {year}
  {1918})}\BibitemShut {NoStop}%
\bibitem [{\citenamefont {{Petrov}}(1954)}]{petrovRussian}%
  \BibitemOpen
  \bibfield  {author} {\bibinfo {author} {\bibfnamefont {A.~Z.}\ \bibnamefont
  {{Petrov}}},\ }\bibfield  {title} {\bibinfo {title}
  {{Классификаця пространств
  определяюшчикх поля тяготения}},\ }\href@noop {}
  {\bibfield  {journal} {\bibinfo  {journal} {Ученые записки
  Казанского государственного
  университета имени В. И. Ульянова-Ленина}\
  }\textbf {\bibinfo {volume} {114}},\ \bibinfo {pages} {55} (\bibinfo {year}
  {1954})}\BibitemShut {NoStop}%
\bibitem [{\citenamefont {Petrov}(2000)}]{Petrov2000}%
  \BibitemOpen
  \bibfield  {author} {\bibinfo {author} {\bibfnamefont {A.~Z.}\ \bibnamefont
  {Petrov}},\ }\bibfield  {title} {\bibinfo {title} {{The Classification of
  Spaces Defining Gravitational Fields}},\ }\href
  {https://doi.org/10.1023/A:1001910908054} {\bibfield  {journal} {\bibinfo
  {journal} {General Relativity and Gravitation}\ }\textbf {\bibinfo {volume}
  {32}},\ \bibinfo {pages} {1665} (\bibinfo {year} {2000})}\BibitemShut
  {NoStop}%
\bibitem [{\citenamefont {{Petrov}}(1969)}]{petrovbook}%
  \BibitemOpen
  \bibfield  {author} {\bibinfo {author} {\bibfnamefont {A.~Z.}\ \bibnamefont
  {{Petrov}}},\ }\href@noop {} {\emph {\bibinfo {title} {Einstein spaces}}}\
  (\bibinfo  {publisher} {Pergamon Press, Oxford},\ \bibinfo {year}
  {1969})\BibitemShut {NoStop}%
\bibitem [{\citenamefont {Pirani}(1957)}]{Pirani1957}%
  \BibitemOpen
  \bibfield  {author} {\bibinfo {author} {\bibfnamefont {F.~A.~E.}\
  \bibnamefont {Pirani}},\ }\bibfield  {title} {\bibinfo {title} {{Invariant
  Formulation of Gravitational Radiation Theory}},\ }\href
  {https://doi.org/10.1103/PhysRev.105.1089} {\bibfield  {journal} {\bibinfo
  {journal} {Phys. Rev.}\ }\textbf {\bibinfo {volume} {105}},\ \bibinfo {pages}
  {1089} (\bibinfo {year} {1957})}\BibitemShut {NoStop}%
\bibitem [{\citenamefont {{Newman}}\ and\ \citenamefont
  {{Penrose}}(1962)}]{npf}%
  \BibitemOpen
  \bibfield  {author} {\bibinfo {author} {\bibfnamefont {E.}~\bibnamefont
  {{Newman}}}\ and\ \bibinfo {author} {\bibfnamefont {R.}~\bibnamefont
  {{Penrose}}},\ }\bibfield  {title} {\bibinfo {title} {{An Approach to
  Gravitational Radiation by a Method of Spin Coefficients}},\ }\href
  {https://doi.org/10.1063/1.1724257} {\bibfield  {journal} {\bibinfo
  {journal} {Journal of Mathematical Physics}\ }\textbf {\bibinfo {volume}
  {3}},\ \bibinfo {pages} {566} (\bibinfo {year} {1962})}\BibitemShut {NoStop}%
\bibitem [{\citenamefont {Penrose}\ and\ \citenamefont
  {Rindler}(1986)}]{Penrose1986}%
  \BibitemOpen
  \bibfield  {author} {\bibinfo {author} {\bibfnamefont {R.}~\bibnamefont
  {Penrose}}\ and\ \bibinfo {author} {\bibfnamefont {W.}~\bibnamefont
  {Rindler}},\ }\href {https://doi.org/10.1017/CBO9780511524486} {\emph
  {\bibinfo {title} {{Spinors and Space-Time}}}},\ \bibinfo {series} {Cambridge
  Monographs on Mathematical Physics}, Vol.~\bibinfo {volume} {2}\ (\bibinfo
  {publisher} {Cambridge University Press},\ \bibinfo {year}
  {1986})\BibitemShut {NoStop}%
\bibitem [{\citenamefont {Visser}(1998{\natexlab{a}})}]{VisserPRL1998}%
  \BibitemOpen
  \bibfield  {author} {\bibinfo {author} {\bibfnamefont {M.}~\bibnamefont
  {Visser}},\ }\bibfield  {title} {\bibinfo {title} {{Hawking Radiation without
  Black Hole Entropy}},\ }\href {https://doi.org/10.1103/PhysRevLett.80.3436}
  {\bibfield  {journal} {\bibinfo  {journal} {Phys. Rev. Lett.}\ }\textbf
  {\bibinfo {volume} {80}},\ \bibinfo {pages} {3436} (\bibinfo {year}
  {1998}{\natexlab{a}})}\BibitemShut {NoStop}%
\bibitem [{\citenamefont {Riemann}(1921)}]{Riemann1921}%
  \BibitemOpen
  \bibfield  {author} {\bibinfo {author} {\bibfnamefont {B.}~\bibnamefont
  {Riemann}},\ }\bibinfo {title} {{{\"U}ber die Hypothesen, welche der
  Geometrie zugrunde liegen}},\ in\ \href
  {https://doi.org/10.1007/978-3-662-24861-4_1} {\emph {\bibinfo {booktitle}
  {{{\"U}ber die Hypothesen, welche der Geometrie zu Grunde liegen}}}},\
  \bibinfo {editor} {edited by\ \bibinfo {editor} {\bibfnamefont
  {H.}~\bibnamefont {Weyl}}}\ (\bibinfo  {publisher} {Springer Berlin
  Heidelberg},\ \bibinfo {year} {1921})\ pp.\ \bibinfo {pages} {1--47},\
  \bibinfo {edition} {2nd}\ ed.\BibitemShut {Stop}%
\bibitem [{\citenamefont {Weinberg}(1972)}]{weinberg1972gravitation}%
  \BibitemOpen
  \bibfield  {author} {\bibinfo {author} {\bibfnamefont {S.}~\bibnamefont
  {Weinberg}},\ }\href@noop {} {\emph {\bibinfo {title} {{Gravitation and
  Cosmology: Principles and Applications of the General Theory of
  Relativity}}}}\ (\bibinfo  {publisher} {Wiley},\ \bibinfo {year}
  {1972})\BibitemShut {NoStop}%
\bibitem [{\citenamefont {Fischer}\ and\ \citenamefont
  {Visser}(2003)}]{FISCHER200322}%
  \BibitemOpen
  \bibfield  {author} {\bibinfo {author} {\bibfnamefont {U.~R.}\ \bibnamefont
  {Fischer}}\ and\ \bibinfo {author} {\bibfnamefont {M.}~\bibnamefont
  {Visser}},\ }\bibfield  {title} {\bibinfo {title} {{On the space-time
  curvature experienced by quasiparticle excitations in the
  Painlev{\'e}–Gullstrand effective geometry}},\ }\href
  {https://doi.org/https://doi.org/10.1016/S0003-4916(03)00011-3} {\bibfield
  {journal} {\bibinfo  {journal} {Annals of Physics}\ }\textbf {\bibinfo
  {volume} {304}},\ \bibinfo {pages} {22} (\bibinfo {year} {2003})}\BibitemShut
  {NoStop}%
\bibitem [{\citenamefont {Fischer}\ and\ \citenamefont {Visser}(2002)}]{iv}%
  \BibitemOpen
  \bibfield  {author} {\bibinfo {author} {\bibfnamefont {U.~R.}\ \bibnamefont
  {Fischer}}\ and\ \bibinfo {author} {\bibfnamefont {M.}~\bibnamefont
  {Visser}},\ }\bibfield  {title} {\bibinfo {title} {{Riemannian Geometry of
  Irrotational Vortex Acoustics}},\ }\href
  {https://doi.org/10.1103/PhysRevLett.88.110201} {\bibfield  {journal}
  {\bibinfo  {journal} {Phys. Rev. Lett.}\ }\textbf {\bibinfo {volume} {88}},\
  \bibinfo {pages} {110201} (\bibinfo {year} {2002})}\BibitemShut {NoStop}%
\bibitem [{\citenamefont {D'Inverno}(1992)}]{d1992introducing}%
  \BibitemOpen
  \bibfield  {author} {\bibinfo {author} {\bibfnamefont {R.}~\bibnamefont
  {D'Inverno}},\ }\href {https://books.google.co.kr/books?id=8nw5fIWhkI4C}
  {\emph {\bibinfo {title} {Introducing Einstein's Relativity}}}\ (\bibinfo
  {publisher} {Clarendon Press, Oxford},\ \bibinfo {year} {1992})\BibitemShut
  {NoStop}%
\bibitem [{\citenamefont {Stephani}\ \emph {et~al.}(2003)\citenamefont
  {Stephani}, \citenamefont {Kramer}, \citenamefont {MacCallum}, \citenamefont
  {Hoenselaers},\ and\ \citenamefont {Herlt}}]{stephani2009exact}%
  \BibitemOpen
  \bibfield  {author} {\bibinfo {author} {\bibfnamefont {H.}~\bibnamefont
  {Stephani}}, \bibinfo {author} {\bibfnamefont {D.}~\bibnamefont {Kramer}},
  \bibinfo {author} {\bibfnamefont {M.}~\bibnamefont {MacCallum}}, \bibinfo
  {author} {\bibfnamefont {C.}~\bibnamefont {Hoenselaers}},\ and\ \bibinfo
  {author} {\bibfnamefont {E.}~\bibnamefont {Herlt}},\ }\href
  {https://doi.org/10.1017/CBO9780511535185} {\emph {\bibinfo {title} {{Exact
  Solutions of Einstein's Field Equations}}}},\ \bibinfo {edition} {2nd}\ ed.,\
  Cambridge Monographs on Mathematical Physics\ (\bibinfo  {publisher}
  {Cambridge University Press},\ \bibinfo {year} {2003})\BibitemShut {NoStop}%
\bibitem [{Note1()}]{Note1}%
  \BibitemOpen
  \bibinfo {note} {We remark that for any tetrad basis variant, orthonormal or
  null, {there are remaining tetrad gauge degrees of freedom}, which can be
  used to bring the tetrad into a suitable form for a specific
  computation.}\BibitemShut {Stop}%
\bibitem [{\citenamefont {Chandrasekhar}(1998)}]{chandrasekharBH}%
  \BibitemOpen
  \bibfield  {author} {\bibinfo {author} {\bibfnamefont {S.}~\bibnamefont
  {Chandrasekhar}},\ }\href {https://books.google.co.kr/books?id=LBOVcrzFfhsC}
  {\emph {\bibinfo {title} {{The Mathematical Theory of Black Holes}}}},\
  International series of monographs on physics\ (\bibinfo  {publisher}
  {Clarendon Press, Oxford},\ \bibinfo {year} {1998})\BibitemShut {NoStop}%
\bibitem [{Note2()}]{Note2}%
  \BibitemOpen
  \bibinfo {note} {{Here, $\protect \mathrm {GL}(n,\protect \mathbb {K})$ is
  the General Linear group associated to a $n$ by $n$ matrix, with $\protect
  \mathbb K$ the mathematical {\protect \em field} ($\protect \mathbb {C}$ for
  the ${\protect \mathcal Q}$ matrix).}}\BibitemShut {Stop}%
\bibitem [{\citenamefont {Moore}(1968)}]{MooreJohn}%
  \BibitemOpen
  \bibfield  {author} {\bibinfo {author} {\bibfnamefont {J.}~\bibnamefont
  {Moore}},\ }\href {https://books.google.co.kr/books?id=Wcw-AAAAIAAJ} {\emph
  {\bibinfo {title} {Elements of Linear Algebra and Matrix Theory}}},\
  International series in pure and applied mathematics\ (\bibinfo  {publisher}
  {McGraw-Hill},\ \bibinfo {year} {1968})\BibitemShut {NoStop}%
\bibitem [{\citenamefont {Shilov}(2012)}]{Shilov}%
  \BibitemOpen
  \bibfield  {author} {\bibinfo {author} {\bibfnamefont {G.~E.}\ \bibnamefont
  {Shilov}},\ }\href {https://books.google.co.kr/books?id=kyHCAgAAQBAJ} {\emph
  {\bibinfo {title} {Linear Algebra}}},\ Dover Books on Mathematics\ (\bibinfo
  {publisher} {Dover Publications},\ \bibinfo {year} {2012})\BibitemShut
  {NoStop}%
\bibitem [{\citenamefont {Kerr}(1963)}]{Kerr1963}%
  \BibitemOpen
  \bibfield  {author} {\bibinfo {author} {\bibfnamefont {R.~P.}\ \bibnamefont
  {Kerr}},\ }\bibfield  {title} {\bibinfo {title} {{Gravitational Field of a
  Spinning Mass as an Example of Algebraically Special Metrics}},\ }\href
  {https://doi.org/10.1103/PhysRevLett.11.237} {\bibfield  {journal} {\bibinfo
  {journal} {Phys. Rev. Lett.}\ }\textbf {\bibinfo {volume} {11}},\ \bibinfo
  {pages} {237} (\bibinfo {year} {1963})}\BibitemShut {NoStop}%
\bibitem [{\citenamefont {Kinnersley}(1969)}]{Kinnersley1969}%
  \BibitemOpen
  \bibfield  {author} {\bibinfo {author} {\bibfnamefont {W.}~\bibnamefont
  {Kinnersley}},\ }\bibfield  {title} {\bibinfo {title} {{Type D Vacuum
  Metrics}},\ }\href {https://doi.org/10.1063/1.1664958} {\bibfield  {journal}
  {\bibinfo  {journal} {Journal of Mathematical Physics}\ }\textbf {\bibinfo
  {volume} {10}},\ \bibinfo {pages} {1195} (\bibinfo {year}
  {1969})}\BibitemShut {NoStop}%
\bibitem [{Note3()}]{Note3}%
  \BibitemOpen
  \bibinfo {note} {{We adopt the notational conventions of Ref.~\cite
  {Datta_2022}, using a subscript 0 in round brackets for background variables
  such as $c_{s(0)}$ which are in general renormalized by the nonlinearity of
  fluid dynamics, while $c_{s0}$ denotes its strictly linearized
  counterpart.}}\BibitemShut {Stop}%
\bibitem [{\citenamefont {Barcel\'o}\ \emph {et~al.}(2003)\citenamefont
  {Barcel\'o}, \citenamefont {Liberati},\ and\ \citenamefont
  {Visser}}]{BLV2003PRA}%
  \BibitemOpen
  \bibfield  {author} {\bibinfo {author} {\bibfnamefont {C.}~\bibnamefont
  {Barcel\'o}}, \bibinfo {author} {\bibfnamefont {S.}~\bibnamefont
  {Liberati}},\ and\ \bibinfo {author} {\bibfnamefont {M.}~\bibnamefont
  {Visser}},\ }\bibfield  {title} {\bibinfo {title} {{Probing semiclassical
  analog gravity in Bose-Einstein condensates with widely tunable
  interactions}},\ }\href {https://doi.org/10.1103/PhysRevA.68.053613}
  {\bibfield  {journal} {\bibinfo  {journal} {Phys. Rev. A}\ }\textbf {\bibinfo
  {volume} {68}},\ \bibinfo {pages} {053613} (\bibinfo {year}
  {2003})}\BibitemShut {NoStop}%
\bibitem [{\citenamefont {Fedichev}\ and\ \citenamefont {Fischer}(2004)}]{CPP}%
  \BibitemOpen
  \bibfield  {author} {\bibinfo {author} {\bibfnamefont {P.~O.}\ \bibnamefont
  {Fedichev}}\ and\ \bibinfo {author} {\bibfnamefont {U.~R.}\ \bibnamefont
  {Fischer}},\ }\bibfield  {title} {\bibinfo {title} {{``Cosmological''
  quasiparticle production in harmonically trapped superfluid gases}},\ }\href
  {https://doi.org/10.1103/PhysRevA.69.033602} {\bibfield  {journal} {\bibinfo
  {journal} {Phys. Rev. A}\ }\textbf {\bibinfo {volume} {69}},\ \bibinfo
  {pages} {033602} (\bibinfo {year} {2004})}\BibitemShut {NoStop}%
\bibitem [{\citenamefont {Riemann}(1860)}]{Riemann1860}%
  \BibitemOpen
  \bibfield  {author} {\bibinfo {author} {\bibfnamefont {B.}~\bibnamefont
  {Riemann}},\ }\bibfield  {title} {\bibinfo {title} {{Ueber die Fortpflanzung
  ebener Luftwellen von endlicher Schwingungsweite}},\ }\href
  {http://eudml.org/doc/135717} {\bibfield  {journal} {\bibinfo  {journal}
  {Abhandlungen der K\"oniglichen Gesellschaft der Wissenschaften in
  G\"ottingen}\ }\textbf {\bibinfo {volume} {8}},\ \bibinfo {pages} {43}
  (\bibinfo {year} {1860})}\BibitemShut {NoStop}%
\bibitem [{\citenamefont {Datta}\ and\ \citenamefont
  {Fischer}(2022)}]{Datta_2022}%
  \BibitemOpen
  \bibfield  {author} {\bibinfo {author} {\bibfnamefont {S.}~\bibnamefont
  {Datta}}\ and\ \bibinfo {author} {\bibfnamefont {U.~R.}\ \bibnamefont
  {Fischer}},\ }\bibfield  {title} {\bibinfo {title} {Analogue gravitational
  field from nonlinear fluid dynamics},\ }\href
  {https://doi.org/10.1088/1361-6382/ac4828} {\bibfield  {journal} {\bibinfo
  {journal} {Classical and Quantum Gravity}\ }\textbf {\bibinfo {volume}
  {39}},\ \bibinfo {pages} {075018} (\bibinfo {year} {2022})}\BibitemShut
  {NoStop}%
\bibitem [{Note4()}]{Note4}%
  \BibitemOpen
  \bibinfo {note} {We note in this regard that the primary constraint on the
  possible Petrov types of a general spatially three-dimensional flow is that
  constant time slices of the analogue metric must be conformally flat in the
  Newtonian lab, as follows from their Painlev\'e-Gullstrand form, in
  distinction to a general solution of the Einstein equations, where no such
  condition applies.}\BibitemShut {Stop}%
\bibitem [{\citenamefont {Visser}\ and\ \citenamefont
  {Weinfurtner}(2005)}]{Visser_2005}%
  \BibitemOpen
  \bibfield  {author} {\bibinfo {author} {\bibfnamefont {M.}~\bibnamefont
  {Visser}}\ and\ \bibinfo {author} {\bibfnamefont {S.}~\bibnamefont
  {Weinfurtner}},\ }\bibfield  {title} {\bibinfo {title} {{Vortex analogue for
  the equatorial geometry of the Kerr black hole}},\ }\href
  {https://doi.org/10.1088/0264-9381/22/12/011} {\bibfield  {journal} {\bibinfo
   {journal} {Classical and Quantum Gravity}\ }\textbf {\bibinfo {volume}
  {22}},\ \bibinfo {pages} {2493} (\bibinfo {year} {2005})}\BibitemShut
  {NoStop}%
\bibitem [{\citenamefont {Visser}(1998{\natexlab{b}})}]{MattCQG}%
  \BibitemOpen
  \bibfield  {author} {\bibinfo {author} {\bibfnamefont {M.}~\bibnamefont
  {Visser}},\ }\bibfield  {title} {\bibinfo {title} {{Acoustic black holes:
  horizons, ergospheres and Hawking radiation}},\ }\href
  {http://stacks.iop.org/0264-9381/15/i=6/a=024} {\bibfield  {journal}
  {\bibinfo  {journal} {Classical and Quantum Gravity}\ }\textbf {\bibinfo
  {volume} {15}},\ \bibinfo {pages} {1767} (\bibinfo {year}
  {1998}{\natexlab{b}})}\BibitemShut {NoStop}%
\bibitem [{\citenamefont {Lense}\ and\ \citenamefont {Thirring}(1918)}]{Lense}%
  \BibitemOpen
  \bibfield  {author} {\bibinfo {author} {\bibfnamefont {J.}~\bibnamefont
  {Lense}}\ and\ \bibinfo {author} {\bibfnamefont {H.}~\bibnamefont
  {Thirring}},\ }\bibfield  {title} {\bibinfo {title} {Über den {Einfluß} der
  {Eigenrotation} der {Zentralkörper} auf die {Bewegung} der {Planeten} und
  {Monde} nach der {Einsteinschen} {Gravitationstheorie}},\ }\href
  {https://ui.adsabs.harvard.edu/abs/1918PhyZ...19..156L} {\bibfield  {journal}
  {\bibinfo  {journal} {Physikalische Zeitschrift}\ }\textbf {\bibinfo {volume}
  {19}},\ \bibinfo {pages} {156} (\bibinfo {year} {1918})}\BibitemShut
  {NoStop}%
\bibitem [{\citenamefont {Baines}\ \emph {et~al.}()\citenamefont {Baines},
  \citenamefont {Berry}, \citenamefont {Simpson},\ and\ \citenamefont
  {Visser}}]{Baines}%
  \BibitemOpen
  \bibfield  {author} {\bibinfo {author} {\bibfnamefont {J.}~\bibnamefont
  {Baines}}, \bibinfo {author} {\bibfnamefont {T.}~\bibnamefont {Berry}},
  \bibinfo {author} {\bibfnamefont {A.}~\bibnamefont {Simpson}},\ and\ \bibinfo
  {author} {\bibfnamefont {M.}~\bibnamefont {Visser}},\ }\bibfield  {title}
  {\bibinfo {title} {{Painlev{\'{e}}-Gullstrand} form of the {Lense-Thirring}
  spacetime},\ }\href@noop {} {\ }\Eprint {https://arxiv.org/abs/2006.14258}
  {arXiv:2006.14258 [gr-qc]} \BibitemShut {NoStop}%
\bibitem [{\citenamefont {Pleba{\'n}ski}(1964)}]{Plebanski}%
  \BibitemOpen
  \bibfield  {author} {\bibinfo {author} {\bibfnamefont {J.}~\bibnamefont
  {Pleba{\'n}ski}},\ }\bibfield  {title} {\bibinfo {title} {{The Algebraic
  Structure of the Tensor of Matter}},\ }\href@noop {} {\bibfield  {journal}
  {\bibinfo  {journal} {Acta Phys. Polon.}\ }\textbf {\bibinfo {volume} {26}},\
  \bibinfo {pages} {963} (\bibinfo {year} {1964})}\BibitemShut {NoStop}%
\bibitem [{\citenamefont {Hall}(1976)}]{Hall}%
  \BibitemOpen
  \bibfield  {author} {\bibinfo {author} {\bibfnamefont {G.~S.}\ \bibnamefont
  {Hall}},\ }\bibfield  {title} {\bibinfo {title} {{The classification of the
  Ricci tensor in general relativity theory}},\ }\href
  {https://doi.org/10.1088/0305-4470/9/4/010} {\bibfield  {journal} {\bibinfo
  {journal} {Journal of Physics A: Mathematical and General}\ }\textbf
  {\bibinfo {volume} {9}},\ \bibinfo {pages} {541} (\bibinfo {year}
  {1976})}\BibitemShut {NoStop}%
\bibitem [{\citenamefont {McIntosh}\ \emph {et~al.}(1981)\citenamefont
  {McIntosh}, \citenamefont {Foyster},\ and\ \citenamefont {Lun}}]{McIntosh}%
  \BibitemOpen
  \bibfield  {author} {\bibinfo {author} {\bibfnamefont {C.~B.~G.}\
  \bibnamefont {McIntosh}}, \bibinfo {author} {\bibfnamefont {J.~M.}\
  \bibnamefont {Foyster}},\ and\ \bibinfo {author} {\bibfnamefont {A.~W.-C.}\
  \bibnamefont {Lun}},\ }\bibfield  {title} {\bibinfo {title} {{The
  classification of the Ricci and Pleba{\'n}ski tensors in general relativity
  using Newman–Penrose formalism}},\ }\href
  {https://doi.org/10.1063/1.524840} {\bibfield  {journal} {\bibinfo  {journal}
  {Journal of Mathematical Physics}\ }\textbf {\bibinfo {volume} {22}},\
  \bibinfo {pages} {2620} (\bibinfo {year} {1981})}\BibitemShut {NoStop}%
\bibitem [{\citenamefont {{Moncrief}}(1980)}]{Moncrief}%
  \BibitemOpen
  \bibfield  {author} {\bibinfo {author} {\bibfnamefont {V.}~\bibnamefont
  {{Moncrief}}},\ }\bibfield  {title} {\bibinfo {title} {{{Stability of
  stationary, spherical accretion onto a Schwarzschild black hole}}},\ }\href
  {https://doi.org/10.1086/157707} {\bibfield  {journal} {\bibinfo  {journal}
  {\apj}\ }\textbf {\bibinfo {volume} {235}},\ \bibinfo {pages} {1038}
  (\bibinfo {year} {1980})}\BibitemShut {NoStop}%
\bibitem [{\citenamefont {Bili{\'c}}(1999)}]{Neven}%
  \BibitemOpen
  \bibfield  {author} {\bibinfo {author} {\bibfnamefont {N.}~\bibnamefont
  {Bili{\'c}}},\ }\bibfield  {title} {\bibinfo {title} {Relativistic acoustic
  geometry},\ }\href {https://doi.org/10.1088/0264-9381/16/12/312} {\bibfield
  {journal} {\bibinfo  {journal} {Classical and Quantum Gravity}\ }\textbf
  {\bibinfo {volume} {16}},\ \bibinfo {pages} {3953} (\bibinfo {year}
  {1999})}\BibitemShut {NoStop}%
\bibitem [{\citenamefont {Visser}\ and\ \citenamefont
  {Molina-Par{\'\i}s}(2010)}]{Molina-Paris}%
  \BibitemOpen
  \bibfield  {author} {\bibinfo {author} {\bibfnamefont {M.}~\bibnamefont
  {Visser}}\ and\ \bibinfo {author} {\bibfnamefont {C.}~\bibnamefont
  {Molina-Par{\'\i}s}},\ }\bibfield  {title} {\bibinfo {title} {Acoustic
  geometry for general relativistic barotropic irrotational fluid flow},\
  }\href {https://doi.org/10.1088/1367-2630/12/9/095014} {\bibfield  {journal}
  {\bibinfo  {journal} {New Journal of Physics}\ }\textbf {\bibinfo {volume}
  {12}},\ \bibinfo {pages} {095014} (\bibinfo {year} {2010})}\BibitemShut
  {NoStop}%
\bibitem [{\citenamefont {Ge}\ \emph {et~al.}(2019)\citenamefont {Ge},
  \citenamefont {Nakahara}, \citenamefont {Sin}, \citenamefont {Tian},\ and\
  \citenamefont {Wu}}]{Nakahara}%
  \BibitemOpen
  \bibfield  {author} {\bibinfo {author} {\bibfnamefont {X.-H.}\ \bibnamefont
  {Ge}}, \bibinfo {author} {\bibfnamefont {M.}~\bibnamefont {Nakahara}},
  \bibinfo {author} {\bibfnamefont {S.-J.}\ \bibnamefont {Sin}}, \bibinfo
  {author} {\bibfnamefont {Y.}~\bibnamefont {Tian}},\ and\ \bibinfo {author}
  {\bibfnamefont {S.-F.}\ \bibnamefont {Wu}},\ }\bibfield  {title} {\bibinfo
  {title} {{Acoustic black holes in curved spacetime and the emergence of
  analogue Minkowski spacetime}},\ }\href
  {https://doi.org/10.1103/PhysRevD.99.104047} {\bibfield  {journal} {\bibinfo
  {journal} {Phys. Rev. D}\ }\textbf {\bibinfo {volume} {99}},\ \bibinfo
  {pages} {104047} (\bibinfo {year} {2019})}\BibitemShut {NoStop}%
\end{thebibliography}%



\end{document}